\documentclass[twocolumn,trackchanges]{aastex631}

\usepackage{soul}
\soulregister{\cite}7 
\soulregister{\citep}7 
\soulregister{\citet}7 
\soulregister{\ref}7 
\soulregister{\pageref}7 

\begin{document}

\title{Two long-period giant planets around two giant stars: HD\,112570 and HD\,154391}

\correspondingauthor{Yu-Juan Liu}
\email{lyj@bao.ac.cn}

\author[0000-0001-6753-4611]{Guang-Yao Xiao}
\affiliation{CAS Key Laboratory of Optical Astronomy, National Astronomical Observatories, Chinese Academy of Sciences, Beijing 100101, China}

\author[0000-0003-3860-6297]{Huan-Yu Teng}
\affiliation{Department of Earth and Planetary Sciences, School of Science, Tokyo Institute of Technology, 2-12-1 Ookayama, Meguro-ku, Tokyo 152-8551, Japan}
\affiliation{CAS Key Laboratory of Optical Astronomy, National Astronomical Observatories, Chinese Academy of Sciences, Beijing 100101, China}

\author[0009-0004-9024-9666]{Jianzhao Zhou}
\affiliation{Institute for Frontiers in Astronomy and Astrophysics, Beijing Normal University, Beijing 102206, China}
\affiliation{Department of Astronomy, Beijing Normal University, Beijing 100875, China}

\author[0000-0001-8033-5633]{Bun'ei Sato}
\affiliation{Department of Earth and Planetary Sciences, School of Science, Tokyo Institute of Technology, 2-12-1 Ookayama, Meguro-ku, Tokyo 152-8551, Japan}

\author[0009-0008-3430-1027]{Yu-Juan Liu}
\affiliation{CAS Key Laboratory of Optical Astronomy, National Astronomical Observatories, Chinese Academy of Sciences, Beijing 100101, China}

\author{Shaolan Bi}
\affiliation{Institute for Frontiers in Astronomy and Astrophysics, Beijing Normal University, Beijing 102206, China}
\affiliation{Department of Astronomy, Beijing Normal University, Beijing 100875, China}

\author{Takuya Takarada}
\affiliation{Astrobiology Center, National Institutes of Natural Sciences, 2-21-1 Osawa, Mitaka, Tokyo 181-8588, Japan}

\author{Masayuki Kuzuhara}
\affiliation{Astrobiology Center, National Institutes of Natural Sciences, 2-21-1 Osawa, Mitaka, Tokyo 181-8588, Japan}

\author[0000-0003-2400-6960]{Marc Hon}  
\affiliation{Institute for Astronomy, University of Hawai'i, 2680 Woodlawn Drive, Honolulu, HI 96822, USA}

\author{Liang Wang}
\affiliation{National Astronomical Observatories, Nanjing Institute of Astronomical Optics $\&$ Technology, Chinese Academy of Sciences, Nanjing 210042, China}
\affiliation{CAS Key Laboratory of Astronomical Optics $\&$ Technology, Nanjing Institute of Astronomical Optics $\&$ Technology, Chinese Academy of Sciences, Nanjing 210042, China}

\author{Masashi Omiya}
\affiliation{Astrobiology Center, National Institutes of Natural Sciences, 2-21-1 Osawa, Mitaka, Tokyo 181-8588, Japan}
\affiliation{National Astronomical Observatory of Japan, National Institutes of Natural Sciences, 2-21-1 Osawa, Mitaka, Tokyo 181-8588, Japan}

\author[0000-0002-7972-0216]{Hiroki Harakawa}
\affiliation{Subaru Telescope, National Astronomical Observatory of Japan, National Institutes of Natural Sciences, 650 North A’ohoku Pl., Hilo, HI, 96720, USA}

\author{Fei Zhao}
\affiliation{CAS Key Laboratory of Optical Astronomy, National Astronomical Observatories, Chinese Academy of Sciences, Beijing 100101, China}

\author{Gang Zhao}
\affiliation{CAS Key Laboratory of Optical Astronomy, National Astronomical Observatories, Chinese Academy of Sciences, Beijing 100101, China}

\author{Eiji Kambe}
\affiliation{Subaru Telescope, National Astronomical Observatory of Japan, National Institutes of Natural Sciences, 650 North A’ohoku Pl., Hilo, HI, 96720, USA}

\author{Hideyuki Izumiura}
\affiliation{Okayama Branch Office, Subaru Telescope, National Astronomical Observatory of Japan, National Institutes of Natural Sciences, Kamogata, Asakuchi, Okayama 719-0232, Japan}

\author{Hiroyasu Ando}
\affiliation{National Astronomical Observatory of Japan, National Institutes of Natural Sciences, 2-21-1 Osawa, Mitaka, Tokyo 181-8588, Japan}

\author{Kunio Noguchi}
\affiliation{National Astronomical Observatory of Japan, National Institutes of Natural Sciences, 2-21-1 Osawa, Mitaka, Tokyo 181-8588, Japan}

\author[0000-0002-9702-4441]{Wei Wang}
\affiliation{CAS Key Laboratory of Optical Astronomy, National Astronomical Observatories, Chinese Academy of Sciences, Beijing 100101, China}

\author{Meng Zhai}
\affiliation{CAS Key Laboratory of Optical Astronomy, National Astronomical Observatories, Chinese Academy of Sciences, Beijing 100101, China}

\author{Nan Song}
\affiliation{China Science and Technology Museum, Beijing 100101, China}

\author{Chengqun Yang}
\affiliation{Shanghai Astronomical Observatory, 80 Nandan Road, Shanghai 200030, China}

\author[0000-0001-6396-2563]{Tanda Li}
\affiliation{Institute for Frontiers in Astronomy and Astrophysics, Beijing Normal University, Beijing 102206, China}
\affiliation{Department of Astronomy, Beijing Normal University, Beijing 100875, China}
\affiliation{School of Physics and Astronomy, University of Birmingham, Edgbaston, Birmingham B15 2TT, UK}

\author[0000-0003-2630-8073]{Timothy D. Brandt}
\affiliation{Department of Physics, University of California, Santa Barbara, Santa Barbara, CA 93106, USA}


\author[0000-0002-9948-1646]{Michitoshi Yoshida}
\affiliation{National Astronomical Observatory of Japan, National Institutes of Natural Sciences, 2-21-1 Osawa, Mitaka, Tokyo 181-8588, Japan}

\author{Yoichi Itoh}
\affiliation{Nishi-Harima Astronomical Observatory, Center for Astronomy, University of Hyogo, 407-2, Nishigaichi, Sayo, Hyogo 679-5313, Japan}

\author{Eiichiro Kokubo}
\affiliation{The Graduate University for Advanced Studies (SOKENDAI), 2-21-1 Osawa, Mitaka, Tokyo 181-8588, Japan}




\begin{abstract}
We present the discoveries of two giant planets orbiting the red giant branch (RGB) star HD\,112570 and the red clump (RC) star HD\,154391, based on the radial velocity (RV) measurements from Xinglong station and Okayama Astrophysical Observatory (OAO). Spectroscopic and asteroseismic analyses suggest that HD\,112570 has a mass of $1.15\pm0.12\,M_{\sun}$, a radius of $9.85\pm0.23\,R_{\sun}$, a metallicity [Fe/H] of $-0.46\pm0.1$ and a ${\rm log}\,\textsl{g}$ of $2.47\pm0.1$. With the joint analysis of RV and Hipparcos-Gaia astrometry, we obtain a dynamical mass of $M_{\rm p}={3.42}_{-0.84}^{+1.4}\ M_{\rm Jup}$, a period of $P={2615}_{-77}^{+85}$ days and a moderate eccentricity of $e={0.20}_{-0.14}^{+0.16}$ for the Jovian planet HD\,112570\,b. 
For HD\,154391, it has a mass of $2.07\pm0.03\,M_{\sun}$, a radius of $8.56\pm0.05\,R_{\sun}$, a metallicity [Fe/H] of $0.07\pm0.1$ and a ${\rm log}\,\textsl{g}$ of $2.86\pm0.1$. The super-Jupiter HD\,154391\,b has a mass of $M_{\rm p}={9.1}_{-1.9}^{+2.8}\ M_{\rm Jup}$, a period of $P={5163}_{-57}^{+60}$ days and an eccentricity of $e={0.20}_{-0.04}^{+0.04}$. We found HD\,154391\,b has one of the longest orbital period among those ever discovered orbiting evolved stars, which may provide a valuable case in our understanding of planetary formation at wider orbits.
Moreover, while a mass gap at $4\,M_{\rm Jup}$ seems to be present in the population of giant stars, there appears to be no significant differences in the distribution of metallicity among giant planets with masses above or below this threshold. Finally, The origin of the abnormal accumulation near 2\,au for planets around large evolved stars ($R_{\star}>21\,R_{\sun}$), remains unclear.
\end{abstract}

\keywords{Exoplanet astronomy(486) --- Evolved stars(481) --- Radial velocity(1332) --- Planet formation(1241)}


\section{Introduction} \label{sec:intro}
To date, more than 5500 exoplanets have been discovered and confirmed through various methods, such as RV, transit, direct imaging, astrometry, and microlensing \citep{Akeson2013}. Most of the planet-hosting stars have spectral types and masses comparable to the Sun, because their abundant absorption lines and relatively small sizes allow for planets to be detected by precision RV and transit surveys. As for intermediate- or higher-mass main sequence (MS) stars ($M_{\star}\gtrsim1.5\,M_{\odot}$), it is difficult to perform such survey due to their larger radius and fewer spectral lines caused by high effective temperature and fast rotation \citep{Lagrange2009}.
However, when those massive stars evolve from the MS phase to the subgiant or giant branch, their radius will expand, and thus the surface temperatures and rotation velocities decline. Therefore, giant stars become feasible targets for RV survey, but the significant jitter from their host stars may limit the detection of smaller worlds like sub-Jovian or super-Earth (e.g., \citealt{Sato2005}). Studies to planets orbiting evolved stars may reveal the effect of stellar evolution on planetary destiny and orbital architecture. 

Since the first exoplanet orbiting a giant star, $\iota$ Draconis b, was detected by the RV technique in 2002 \citep{Frink2002}, several RV surveys aiming to unveil planets around giant stars have been carried out by different scientific teams, including the Lick giants survey \citep{Frink2001,Reffert2015}, the ESO planet search program \citep{Setiawan2003}, the Okayama Planet Search Program (OPSP; \citealt{Sato2005}), the Tautenburg Observatory Planet Search \citep{Hatzes2005, Dollinger2007}, the ``Retired A Stars and Their Companions'' \citep{Johnson2007}, the Penn State-Toru{\'n} Planet Search \citep{Niedzielski2007}, the Bohyunsan Optical Astronomy Observatory (BOAO; \citealt{Han2010}) k-giant survey, the Pan-Pacific Planet Search (PPPS; \citealt{Wittenmyer2011ApJ}), and the EXoPlanet aRound Evolved StarS project (EXPRESS: \citealt{Jones2011}). Up to now, about 200 planets around evolved stars are listed in NASA Exoplanet Archive \citep{Akeson2013}, and almost all of them belong to gas giant planets or brown dwarfs. Some properties of those planetary systems have been widely investigated in recent works.

Hot or warm Jupiters were often detected orbiting dwarf stars, and were previously thought to be unlikely to survive in close-in orbit when orbiting evolved stars \citep{Jones2014}. Theoretical studies suggested that under the influence of tidal interactions, close-in planets will spiral inwards and eventually be engulfed by their host during expansion on the giant phase (e.g. \citealt{Kunitomo2011, Villaver2009}). However, both RV and transit surveys have discovered several short-period giant planets. For example, HD\,167768\,b, a Jovian planet around a deeply evolved star ($1.08\,M_{\odot}$, $9.7\,R_{\odot}$ and ${\rm log}\,g=2.5$) in a 20.65 days and eccentric orbit  \citep{Teng2023}.
These close-in planets can provide excellent insight for studying star-planet interaction and planetary orbit migration. Recently, some works point out that transiting giant planets around giant stars prefer more eccentric orbits than those around MS stars \citep{Jones2018, Grunblatt2018, Grunblatt2022}. \citet{Grunblatt2022} reported a loglinear trend in the period-eccentricity plane for planets transiting evolved stars, which might be explained by the enhancement of tidal interaction or planet-planet scattering. However, the eccentricity of the long-period planetary populations ($>0.1\,{\rm au}$) revealed by RV method tend to have lower values than planets around MS dwarfs \citep{Jones2014}. This difference in eccentricity distribution between short- and long-period planets may imply that post-MS systems are relatively dynamically inactive at larger orbital separations.

The dependence of giant planet occurrence on stellar mass and metallicity has been explored in the past two decades. The occurrence was found to approximately increase with stellar masses below $\sim2\,M_{\odot}$ \citep{Johnson2010, Ghezzi2018, Bowler2010} and decrease beyond $\sim2\,M_{\odot}$ \citep{Jones2016}. More recently, \citet{Wolthoff2022} identified a peak of $1.68\pm0.59\,M_{\odot}$ in the giant planet occurrence rate with respect to stellar mass.
Meanwhile, the direct imaging surveys uncovered a strong correlation that compared with solar-mass FGK stars, more massive stars ($M_{\star}\gtrsim1.5\,M_{\odot}$) exhibit higher occurrence of substellar companions on wide separation ($\rm \gtrsim5\,au$) \citep{Nielsen2019,Vigan2021}. 
In contrast, the transit surveys found a lower frequency for close-in giant planets around intermediate-mass stars \citep{Zhou2019, Beleznay2022, Sebastian2022}. 
These results raise the question of why massive stars possess less close-in planet but more planets at large orbital distance than solar-like stars. 
One possible explanation is linked to the relatively short disk lifetime of massive stars, i.e., the disk lifetime generally decrease with increasing stellar mass (e.g., \citealt{Ribas2015, Komaki2021, Ronco2023arXiv}), and the fast dispersal of a disk by photoevaporation (e.g., \citealt{Owen2012,Kunitomo2021}) results in the halt of the gas-driven inward migration of giant planets. 
However, the disk evolutionary pathways for massive stars remain to be explored.

As for the dependence on metallicity, the well-known giant planet-metallicity correlation \citep{Fischer2005} was proposed for planets around MS stars, i.e., giant planets form preferentially around metal-rich stars, whose disks harbor more solids or planetary building materials, and less frequently around metal-poor host stars \citep{Gonzalez1997, Santos2001,Fischer2005,Mordasini2012}. However, this correlation seems to be controversial for planet-hosting giant star systems. Some studies found no evidence for such correlation \citep{Takeda2008,Maldonado2012}, implying the disk instability scenario of planet formation, while other works reported a positive planet-metallicity correlation \citep{Reffert2015,Jones2016,Wittenmyer2017, Wolthoff2022}, supporting the core-accretion paradigm (e.g., \citealt{Pollack1996, Ida2004,Santos2004}).
Given that the small sample and selection criteria might have a significant influence on determining the true properties of evolved systems, it is too premature for us to make a defined conclusion \citep{Lee2012, Dollinger2021}.

One of the most notable caveats for RV planetary hunting around evolved stars is that the stellar intrinsic variability can masquerade as a planet. For example, \citet{Hatzes2015} reported a long-lived RV-signal ($\sim 629$ day) for K giant star Aldebaran, and interpreted it as caused by a massive Jovian, but subsequent analyses by \citet{Reichert2019} challenged the planet hypothesis. They proposed that oscillatory convective modes \citep{Saio2015} might be a plausible alternative explanation of the observed RV variations. Another typical case is $\gamma\,{\rm Dra}$. \citet{Hatzes2018} found its RV variations surprisingly disappeared for 3 years and returned with a noticeable phase shift. These cases warn us to carefully interpret the periodic RV-signal from giant stars. Apart from continuous RV monitoring, there are additional tests that allow to prove an RV-signal is due to a planet. 
One method is combining RV and astrometric measurements. Astrometry, especially in Gaia era, can be used to corroborate a signal in another dimension \citep{Hill2021}. 
Another method is to obtain additionally RV measurements at infrared wavelength (e.g., \citealt{Trifonov2015}). The RV signals should be consistent in both optical and infrared domains if they are indeed caused by bona fide planets. 

In comparison with MS stars, the mass of evolved stars is difficult to determine owing to dense evolutionary tracks of different masses occupying almost identical areas in Hertzsprung-Russell (HR) diagram. Fortunately, with the vast releases of photometric data obtained from Kepler mission \citep{Borucki2010} and Transiting Exoplanet Survey Satellite (TESS; \citealt{Ricker2015}), asteroseismology, the study of solar-like oscillations, can be a supplemental and powerful approach to precisely measure the fundamental parameters of those giant stars (e.g., \citealt{Huber2010,Stello2013,Stello2022,Hon2022}). 

This paper is organised as follows. In Section 2, we describe the observations of HD\,112570 and HD\,154391, including photometry of TESS, spectroscopy and astrometry. Section 3 and 4 present the detailed analyses of stellar properties and planetary orbital solutions, respectively. The analyses of line profile and chromospheric activity are presented in Section 5. In Section 6, we discuss the lithium abundances and planetary formation about two systems, and make a glance at the strange overabundance of planets around large giant stars.
Finally, we give a brief summary about this paper in Section 7.

\section{Observations} \label{sec:obser}

\subsection{TESS Photometry} \label{subsec:tess}

The Transiting Exoplanet Survey Satellite (TESS), launched in 2018, has the primary objective of discovering and characterizing exoplanets \citep{Ricker2015}. By covering 13 sectors per half year, TESS provides high-precision photometric data for a wide range of bright stars across the sky. Additionally, it acquires full-frame images (FFIs) every 30 minutes as part of its extended mission, with a 2-minute and 20-second cadence for target observations and FFIs obtained every 10 minutes. For our study, we downloaded the 2-minute cadence light curves (PDCSAP data) of our selected targets from the Mikulski Archive for Space Telescopes (MAST), which were processed by the TESS Science Processing Operations Center (SPOC) pipeline \citep{Jenkins2017, Twicken2016}.

Since the PDCSAP data underwent systematic flux removal for each sector, we only applied a $5\,\sigma$ clipping method to remove outliers and normalized each sector by dividing it by the median value.
In Appendix (Figure \ref{fig:twotargetlc}), we present the processing results for HD\,112570 (left panel) and HD\,154391 (right panel), encompassing four and nineteen observed sectors, respectively. The presence of a gap midway through each sector is attributed to the data downlink, resulting in a temporal separation of the two spacecraft orbits. Although sector 49 of HD\,112570 exhibited a higher noise level, this noise did not adversely affect our asteroseismic analysis, and thus, we retained the data for further examination.

\subsection{Radial Velocities} 

\subsubsection{OAO Observations} \label{subsec:oao}

The first spectrum of the stars was obtained in 2005 May using the 1.88 m reflector with HIgh Dispersion Echelle
Spectrograph (HIDES; \citealt{Izumiura1999}) at Okayama Astrophysical Observatory (OAO). An iodine absorption cell was placed in the HIDES optical path in order to provide wavelength reference in the range of $5000\sim5800$ \AA. The initial coverage of HIDES was designed to $5000\sim6100$ \AA\, with a single 2 K $\times$ 4 K CCD. In 2007 December, the upgrade CCD mosaic of three significantly widened the wavelength region to $3700\sim7500$ \AA, and enabled the simultaneous measurements of stellar activities and line profiles as well as RVs. 
The slit width was set to 200 $\mu \rm m$ (0.\arcsec76) corresponding to a resolution of $R=\lambda/\Delta\lambda=67,000$ by about 3.3-pixel sampling.

In 2010, a new high-efficiency fiber-link system with its own iodine cell was available \citep{Kambe2013}. In 2018, the optical path of the fiber-link system was re-arranged and the optical instruments were placed on a new stablized platform in the precise temperature-controlled Coud{\'e} room. The width of 1.\arcsec05 of the sliced image corresponded to a resolution of 55,000 by 3.8-pixel sampling.

Observations of two stars were conducted by both the conventional slit mode (hereafter HIDES-S) and the fiber mode (pre-upgrade in 2018: HIDES-F1 and post-upgrade in 2018: HIDES-F2). 
The reduction of echelle data was performed using IRAF\footnote{IRAF is distributed by the National Optical Astronomy Observatories, which is operated by the Association of Universities for Research in Astronomy, Inc. under a cooperative agreement with the National Science Foundation, USA} software in a standard way (i.e., bias subtraction, flat fielding, scattered light subtraction, and spectrum extraction).
Due to severe aperture overlaps among $3700\sim4000$ \AA\, in fiber mode, the scatter light could not be removed with IRAF, and we therefore discarded the overlapped apertures which caused loss of the Ca {\uppercase\expandafter{\romannumeral2}} H lines.

RV analysis for HIDES data was performed by the method of \citet{Sato2002,Sato2012} based on \citet{Butler1996}. We modeled the $\rm{I_{2}}$-superposed spectra ($5000\sim5800$ \AA) by using the stellar templates which were extracted by deconvolving pure stellar spectra with an instrumental profile determined from $\rm{I_{2}}$--superposed B-type star or flat spectra. The final RV values and uncertainties were derived from the average of the measurements of hundreds of segments with a typical width of 150 pixels.

\subsubsection{Xinglong Observations} \label{subsec:xl}
The Xinglong Planet Search Program started in 2005 within a
framework of international collaborations between China and Japan aiming to probe exoplanets around intermediate-mass G-type (and early K-type) giant stars.
The earliest observation of two stars at Xinglong Station began in 2005 May using 2.16 m reflector and the Coud{\'e} Echelle Spectrograph (CES; \citealt{Zhao2001}). An iodine cell was installed in front of the entrance slit of the spectrograph, which provides a fiducial wavelength reference for precise RV measurements.
The single 1 K $\times$ 1 K CCD (pixel size of 24 $\times$ 24 $\mu \rm m^2$; hereafter CES-O) covered a wavelength range from 3900 to 7260 \AA\, with
a spectral resolution of 40,000 by 2-pixel sampling. Owing to the small format of CCD, only a narrow waveband of $\Delta \lambda\sim470$ \AA\, was selected for RV measurements. RV analysis for CES data was
performed by the optimized method of \citet{Sato2002}. Five Gaussian profiles are used to reconstruct the instrumental profile, and a second-degree Legendre polynomial is used to describe the wavelength scale \citep{Liu2008}. The available spectrum is divided into about 40 segments, and similar to OAO, the final RV values and uncertainties are derived from the average of measurements in each segment.

In 2009 March, a new 2 K $\times$ 2 K CCD (hereafter CES-N) with smaller pixel size of 13 $\times$ 13 $\mu \rm m^2$ was used to replace the old one, which slightly improved the RV precision but the wavelength coverage was unchanged \citep{Wang2012}. Since 2012 June, the observations were conducted with the newly developed High Resolution Spectrograph (HRS) attached at the Cassegrain focus of the 2.16 m telescope. As the successor of the CES, the fiber-fed HRS can provide a higher wavelength resolution and optical throughput. The new 4 K $\times$ 4 K CCD simultaneously covered a wider wavelength of $3700\sim9200$ \AA, and the slit width was set to 190 $\mu \rm m$ which leads to a resolution of 45,000 by 3.2-pixel sampling. RV analysis for HRS data was
performed by the optimized method of \citet{Sato2002,Sato2012}.

\subsection{Absolute Astrometry}
Absolute astrometry of two stars is from a cross-calibrated Hipparcos-Gaia Catalog of Accelerations (HGCA: \citealt{Brandt2018, Brandt2021}), mainly consisting of parallax ($\varpi$), position ($\alpha,\ \delta$) and three proper motions ($\mu_{\alpha},\ \mu_{\delta}$). 
Since Hipparcos \citep{Perryman1997, vanLeeuwen2007} and Gaia EDR3 \citep{GaiaCollaboration2016,GaiaCollaboration2021,Lindegren2021} have a temporal baseline of $\sim25$ years, the proper motion anomalies might indicate the acceleration (e.g. \citealt{Kervella2022}) of a star which could be caused by an invisible companion. Furthermore, the difference in positions between two measurements can provide an additional proper motion scaled by the temporal baseline. In HGCA, the reference frame of two satellites has been placed in a common inertial frame, and the systematic error has also been calibrated \citep{Brandt2018}. Therefore, we directly use the absolute astrometry from HGCA (EDR3 version, \citealt{Brandt2021}) to perform joint analysis with RVs.
The detailed astrometry of two stars can be found in Table \ref{Tab:hgca}.
\section{Stellar properties} \label{sec:stell}

\begin{figure}[ht!]
\plotone{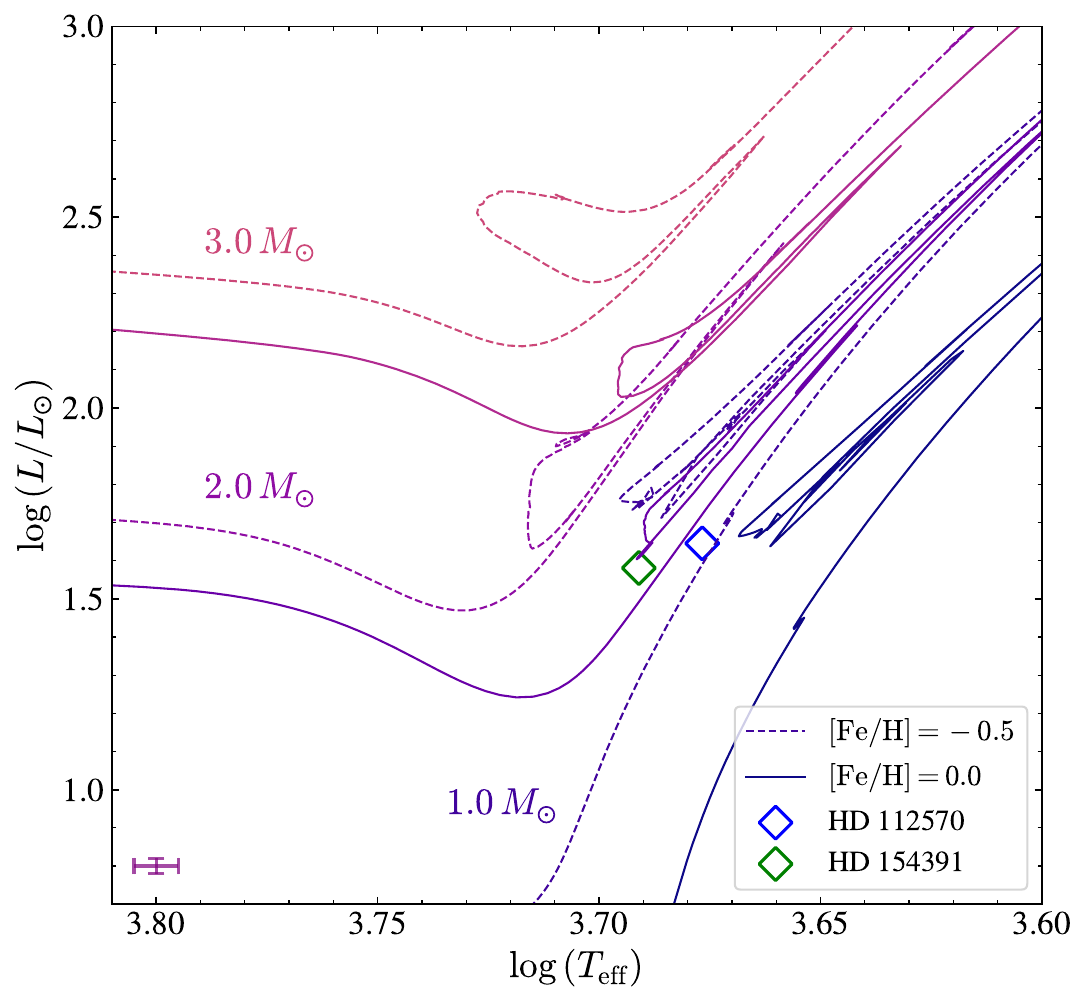}
\caption{HR diagram for two stars. We illustrate MIST evolutionary track of $1\,M_{\sun}$, $2\,M_{\sun}$ and $3\,M_{\sun}$ stars. The metallicity of 0.0 and $-0.5$ dex are shown by solid and dashed line, respectively. Two stars are marked by blue and green open diamond, respectively. The typical error bar is shown in the bottom left part. Both stars have significantly evolved off the MS phase.
\label{fig:hr_dia}}
\end{figure}

\begin{figure}[ht!]
\plotone{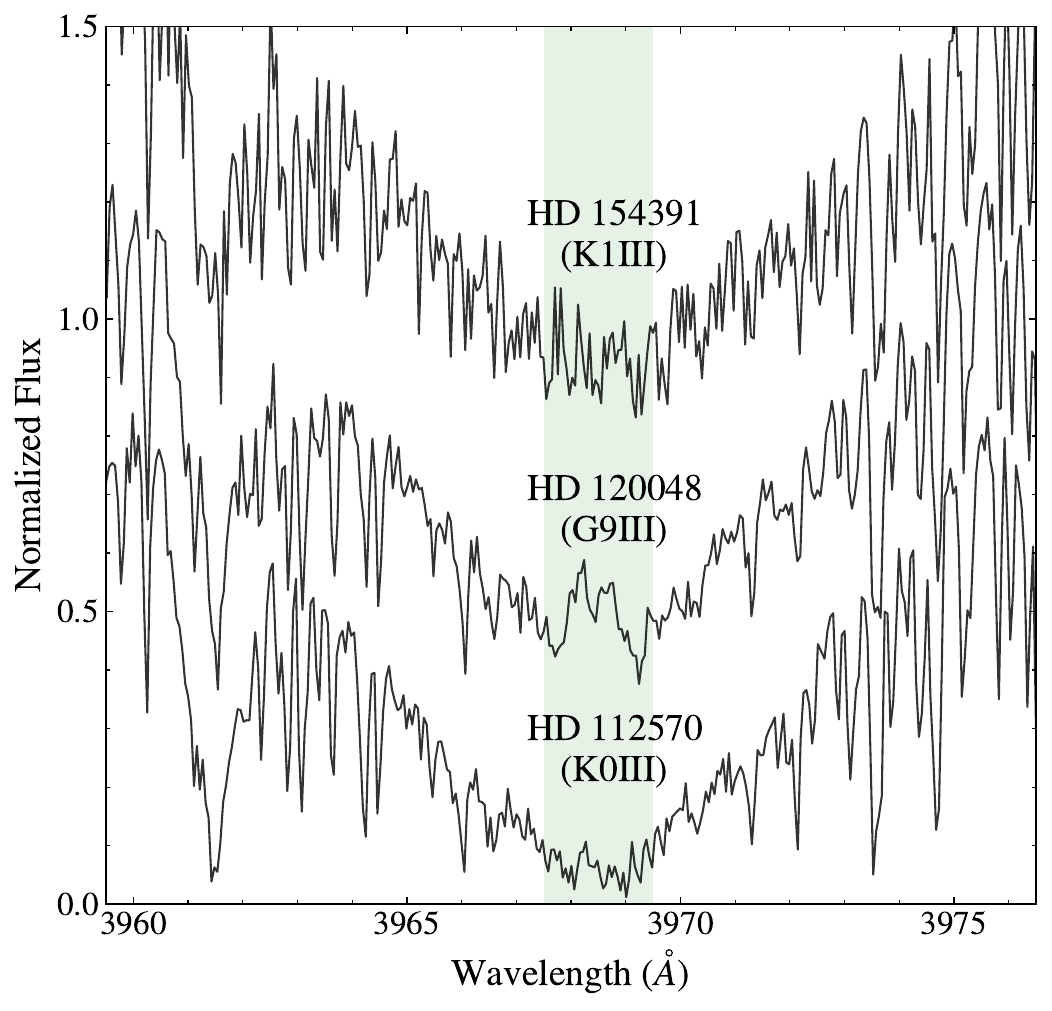}
\caption{Spectra in the region of Ca \uppercase\expandafter{\romannumeral2} H lines. The chromospherically active star HD\,120048 are also shown as a comparison.
\label{fig:caiih}}
\end{figure}

\subsection{Spectroscopy} \label{subsec:spect}
The stellar atmospheric parameters (namely, effective temperature $T_{\rm eff}$, Fe abundance ${\rm [Fe/H]}$ and surface gravity ${\rm log}\,\textsl{g}$) are determined by measuring  iron equivalent widths of $\rm I_2$-free spectra combined with a atmospheric model, following the methodology of \citet{Liu2008}.
For HD\,112570, we obtain $T_{\rm eff}=4672\pm100$\,K, $\rm [Fe/H]=-0.46\pm0.1$\,dex and ${\rm log}\,g=2.47\pm0.1$\,cgs, consistent with that of \citet{Liu2010}. 
For HD\,154391, we obtain $T_{\rm eff}=4807\pm100$\,K, $\rm [Fe/H]=0.07\pm0.1$\,dex and ${\rm log}\,g=2.86\pm0.1$\,cgs, comparable with the values derived by \citet{Tautvais2020}.
Moreover, we perform a global fit of stellar parameters (e.g., stellar mass $M_{\star}$, radius $R_{\star}$ and age $t$) using \texttt{isochrones} \citep{Morton2015} together with the Gaia magnitudes $GG_{\rm BP}G_{\rm RP}$, the Tycho-2 magnitudes $B_{T}V_{T}$, the 2MASS magnitudes $JHK_{s}$, the Gaia DR3 parallax, and asteroseismic parameters (see next section). The detailed stellar parameters are listed in Table \ref{tab:example}.  
Figure \ref{fig:hr_dia} plots the location of two giant stars in HR diagram with evolutionary tracks downloaded from MESA Isochrones \& Stellar Tracks (MIST, \citealt{Paxton2011, Choi2016, Dotter2016}).
As can be seen, HD\,112570 still resides in the first ascent phase, while HD\,154391 has already reached the horizontal branch. 
The post-MS stars might have substantial mass-loss, indicating the stellar masses derived here may differ from the ones on the MS stage. However, following the Equation (5) of \citet{Kunitomo2011} adopted from \citet{Reimers1975}, we find the mass-loss for the less-evolved RGB star, HD\,112570, and for the high-mass RC star, HD\,154391, are not significant ($\lesssim1\%$).
Figure \ref{fig:caiih} shows the spectra of Ca \uppercase\expandafter{\romannumeral2} H line region. Comparing with chromospherically active star HD\,120048, we find no significant emission in the core of Ca \uppercase\expandafter{\romannumeral2} H line for two stars, indicating a quiet chromosphere.  

\begin{table}

\caption{Stellar Parameters}
\label{tab:example}
\begin{tabular}{lccc}
\hline\hline
Parameter & HD\,112570&HD\,154391 & Source \\
\hline
\multicolumn{4}{c}{Basic Properties}\\
HIP ID & 63211&83289 & 1\\
TIC &  137004295 & 424731682  & 2\\
TESS Mag. & 5.071 & 5.758 & 2\\
Sp. Type & K0 \uppercase\expandafter{\romannumeral3}& K1 \uppercase\expandafter{\romannumeral3} & 3, 4\\
$\varpi$ (mas)& $9.948\pm0.027$ & $9.844\pm0.022$ &5\\
$V$ & $6.110\pm0.009$ & $6.143\pm0.010$  & 6\\
$B - V$ & $1.020\pm0.011$ & $1.021\pm0.018$ & 6 \\
\multicolumn{4}{c}{Spectroscopy}\\
$T_{\rm eff}\,(\rm K)$ &$4672\pm100$&$4807\pm100$ & 8\\
${\rm log}\,\textsl{g}$ (cgs) &$2.47\pm0.1$&$2.86\pm0.1$ & 8\\
$\rm [Fe/H]$ (dex) &$-0.46\pm0.1$&$0.07\pm0.1$ & 8\\
\multicolumn{4}{c}{Asteroseismology}\\
$\Delta\nu$ ($\mu{\rm Hz}$) &$4.70\pm0.11$&$7.79\pm0.04$ & 8\\
$\Delta\nu$ ($\mu{\rm Hz}$) &-&$7.82\pm0.03$ & 7\\ 
$\nu_{\rm max}$ ($\mu{\rm Hz}$) &$42.80\pm4.25$&$94.31\pm1.46$ & 8\\
$\nu_{\rm max}$ ($\mu{\rm Hz}$) &-&$93.96\pm0.98$& 7\\
Phase$^{\rm a}$ & RGB &CHeB& 7, 8\\
$M_{\star}\,(M_{\sun})^{\rm b}$  & $1.15\pm0.27$ & $2.00\pm0.11$ & 8\\
$R_{\star}\,(R_{\sun})^{\rm b}$  & $9.61\pm0.61$ & $8.46\pm0.13$& 8\\
${\rm log}\,\textsl{g}$ (cgs)$^{\rm b}$ & $2.53\pm0.01$ & $2.88\pm0.01$ &  8\\
\multicolumn{4}{c}{Isochrones}\\
$T_{\rm eff}\,(\rm K)$ &$4750\pm51$& $4909\pm32$ & 8\\
${\rm log}\,\textsl{g}$ (cgs) &$2.51\pm0.05$&$2.89\pm0.01$ & 8\\
$\rm [Fe/H]$ (dex) &$-0.40\pm0.07$&$0.10\pm0.05$ & 8\\
$L_{\star}\,(L_{\sun})$ &$44.32\pm3.37$&$38.12\pm0.99$ & 8\\
$M_{\star}\,(M_{\sun})$  & $1.15\pm0.12$ &$2.07\pm0.03$ & 8\\
$R_{\star}\,(R_{\sun})$  & $9.85\pm0.23$ &$8.56\pm0.05$ & 8\\
${\rm Age} $ (Gyr)  &$5.45\pm2.06$ &$1.10\pm0.04$ & 8\\

\hline
\end{tabular}
\tablecomments{$^{\rm a}$Evolutionary state: RGB = Hydrogen-shell burning giant star, CHeB = Core helium-burning giant star.\\
$^{\rm b}$Derived from asteroseismic scaling relations.
}
\tablerefs{(1) \citet{ESA1997}, (2) \citet{Stassun2018}, (3) \citet{Upgren1962}, (4) \citet{Halliday1955}, (5) \citet{GaiaCollaboration2021}, (6) \citet{Perryman1997}, (7) \citet{Hon2022}, (8) this work}
\end{table}

\subsection{Asteroseismology} \label{subsec:aster}

\begin{figure*}[ht]
\centering
\includegraphics[width=0.48\linewidth]{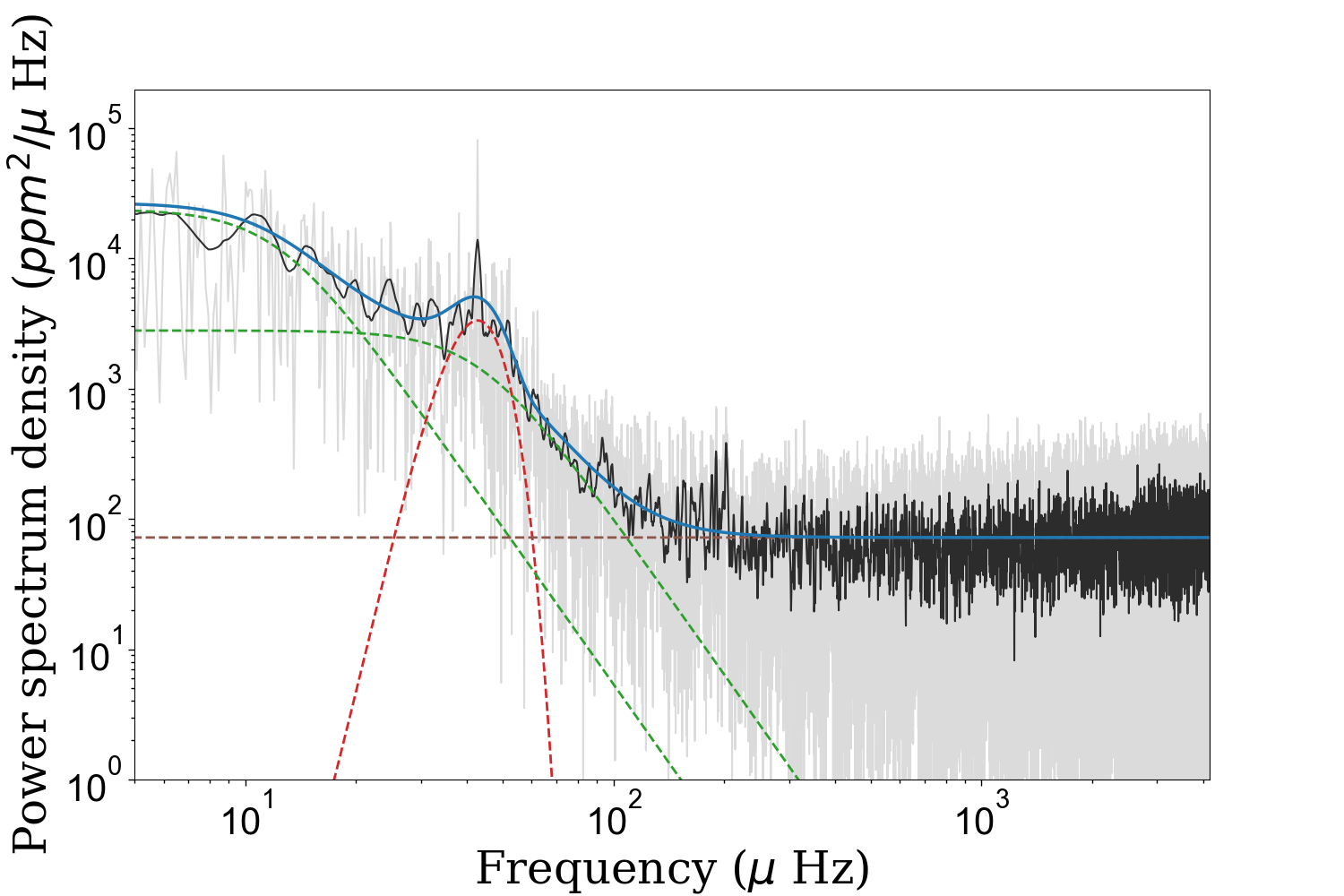}
\includegraphics[width=0.48\linewidth]{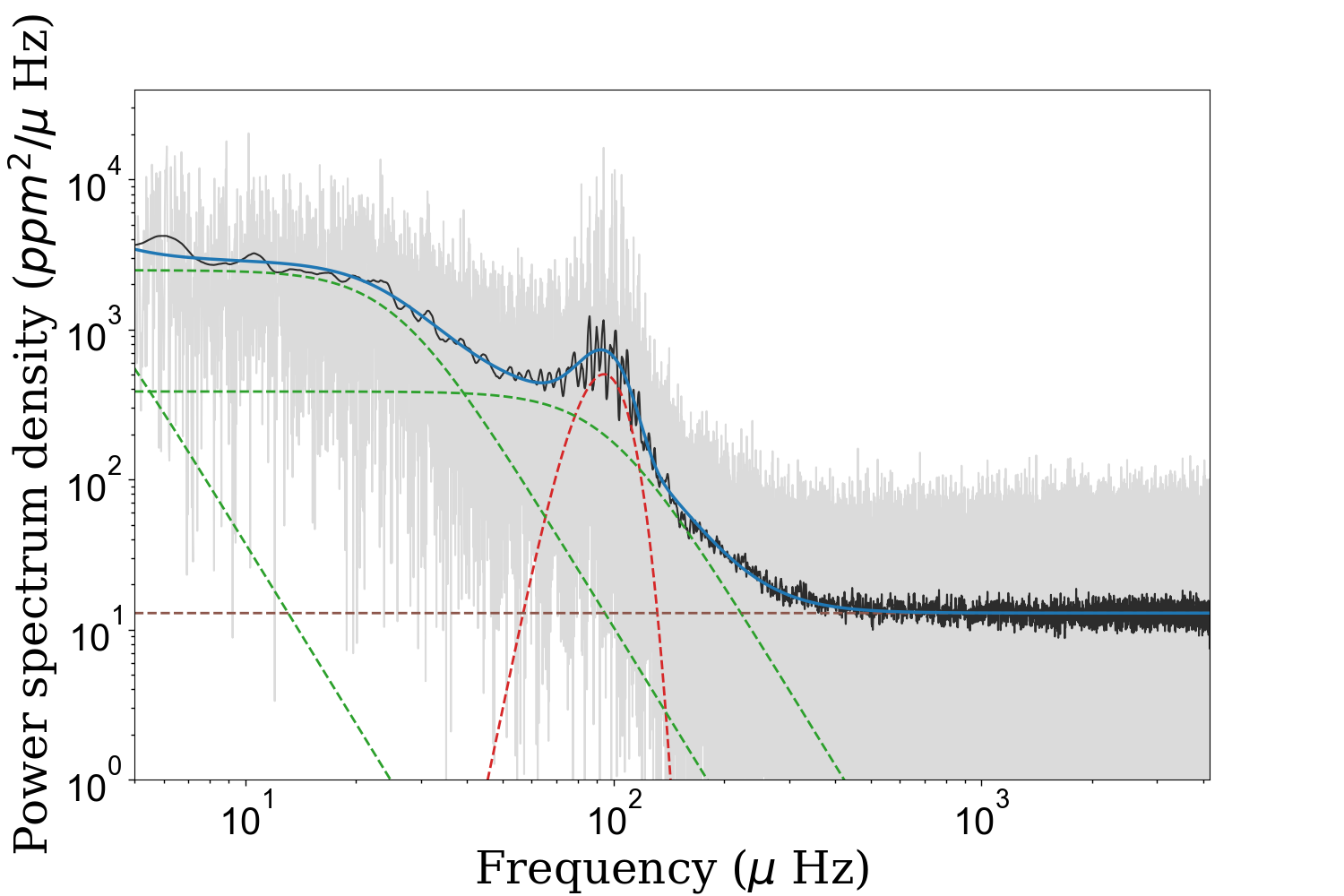}
\caption{Fitting results of HD\,112570 (left panel) and HD\,154391 (right panel). In each panel, the gray line depicts the real data, and the black line shows the data smoothed using a $3\mu$Hz window.  The solid blue line represents the result of MCMC fitting, while the red dashed line corresponds to the fitted Gaussian envelope.  Additionally, the green dashed curve corresponds to the three Harvey components, and the brown straight line represents the white noise.\label{fig:fit_mcmc_targets}}
\end{figure*}

We applied the Lomb-Scargle Periodograms method to TESS photometric data  for power density spectrum generation \citep[e.g.,][]{VanderPlas2018}. 
The power density spectrum was then fitted using the Maximum Likelihood Estimate (MLE) method with a model comprising a Gaussian envelope, three background Harvey components, and white noise, following \citet{Huber2009a} and \citet{Chontos2022} (SYD pipeline). 
For refined estimates, we used a Bayesian approach with Markov-Chain Monte-Carlo simulation \citep{zinn2019,Themessl2020} (Figure \ref{fig:fit_mcmc_targets}), initializing with MLE results.

We employed the autocorrelation function (ACF) method to measure $\Delta\nu$ values, as shown in Figure \ref{fig:Deltanu_targets}. Formal uncertainties for $\Delta\nu$ were obtained by subjecting the power-density spectrum to 500 perturbations, employing a $\chi^2$ distribution with two degrees of freedom. We repeated the fitting procedure for each perturbation and quantified the standard deviation of the output parameter distributions as the formal uncertainty \citep{Huber2011}.

We utilized the following scaling relations to determine the radius ($R$) and mass ($M$):

\begin{equation}
    \frac{M}{M_{\sun}} \approx \left(\frac{\nu_{\rm max}}{\nu_{\rm max,\sun}}\right)^3 \left( \frac{\Delta\nu}{f_{\Delta\nu}\Delta\nu_{\sun}}\right)^{-4} \left(\frac{T_{\text{eff}}}{T_{\text{eff},\sun}}\right)^{3/2},\\
    \label{eq:M_sca}
\end{equation}
\begin{equation}
    \frac{R}{R_{\sun}} \approx \left(\frac{\nu_{\rm max}}{\nu_{\rm max,\sun}}\right) \left( \frac{\Delta\nu}{f_{\Delta\nu}\Delta\nu_{\sun}}\right)^{-2} \left(\frac{T_{\text{eff}}}{T_{\text{eff},\sun}}\right)^{1/2}.
    \label{eq:R_sca}
\end{equation}

where $\nu_{\rm max}$ is expected to be related to stellar fundamental properties, scaling with the acoustic cutoff frequency \citep{Brown1991,Kjeldsen-Bedding1995,Belkacem2011}. Meanwhile, the $\Delta \nu$, which probes the sound speed profile, is presumed to be related to the square root of the mean stellar density \citep{Ulrich1986}. Correction factors $f_{\Delta\nu}$ are used to adjust for any potential deviations from the expected scales.
The solar reference values of $\nu_{\rm max, \sun} = 3090$ $\mu$Hz, $\Delta\nu_{\sun} = 135.1$ $\mu$Hz, and $T_{\rm eff, \sun} = 5777$ K adopted in this study are from \citet{Huber2011,Huber2013}.

Many studies have shown that correction factors can improve the consistency of independently measured fundamental parameters from various sources, such as eclipsing binaries and optical interferometry \citep{Gaulme2016,Sharma2016,Brogaard2018,Huber2017a,Hall2019,Zinn2019b}. 
The scaling relations exhibit an explicit temperature dependence, and the correction term $f_{\Delta\nu}$ is commonly determined with respect to stellar models, considering factors such as mass, temperature, metallicity, and evolutionary phase of the star \citep{Sharma2016,Rodrigues2017,Serenelli2017,Liyg2021}. For estimates of masses and radii, we adopt the correction method proposed by \citet{Sharma2016}, and the evolutionary phase is determined by deep learning classifier \citep{Hon2017,Hon2018}, which uses the frequency distribution of oscillation modes within a star’s collapsed echelle diagram (e.g., the top panel figure in Figure \ref{fig:Deltanu_targets}). The scaling relations reveal a mass of $1.15\pm0.27\,M_{\sun}$ for HD\,112570, and a mass of $2.00\pm0.11\,M_{\sun}$ for HD\,154391, respectively. 
In order to minimize the mass uncertainties of the planet propagated by host mass, we utilize the stellar mass inferred from the global fit of \texttt{isochrones} as the prior in our subsequent orbital analysis.

\section{Orbital fit and planetary parameters} \label{sec:orbit}
Since the RV time series obtained from Xinglong observation contain more than one measurement on some night, we firstly bin the data each night in order to eliminate the high-frequency signal. Besides, the RV offsets between HIDES-S, -F1 and -F2 are fixed to 0 for using the same reference spectra of each mode, which makes it conductive to reduce free parameters and facilitates orbital fit. 
However, the accuracy of planetary parameters inevitably declines. 
\citet{Teng2023arXiv} examined the 20-yr long-term stability of HIDES by re-extracting the RVs of $\tau\,$Cet with the same reference spectrum.
They found an offset between HIDES-S and HIDES-F1, -F2 at a maximum of $\sim5\,\rm m\,s^{-1}$, and an offset between HIDES-F1 and -F2 of $\sim2.5\,\rm m\,s^{-1}$.
In this study, we use the same method as that of \citet{Teng2023arXiv} by shifting the RV offsets between individual modes based on the mean levels instead of setting them as free parameters.
We refer the reader to Appendix 1 of \citet{Teng2023arXiv} for more details of instrumental stability of HIDES.

We apply generalized Lomb-Scargle periodogram (GLS: \citealt{Zechmeister2009}) to search periodic signal in the RV time series. The significance of the periodicity is assessed by means of False Alarm Probability (FAP).

The RV data are initially fit using Keplerian orbital modeling toolkit \texttt{RadVel} \citep{Fulton2018}. The fitting bases are selected as orbital period $P$, time of inferior conjunction $T_{c}$, the combining form of eccentricity $e$ and argument of periastron $\omega$ (i.e, $\sqrt{e}\ {\rm sin}\ \omega$, $\sqrt{e}\ {\rm cos}\ \omega$), and the logarithm of RV semi-amplitude ${\rm log}\,K$. In addition, the RV jitter $\sigma$ and RV offset $\gamma$ are also set as free parameters. The priors for each parameter are listed in Table \ref{Tab:prior}. 

\begin{table}
\centering
\caption{Adopted Prior for \texttt{RadVel}}\label{Tab:prior}
\begin{tabular}{lc}
\hline \hline
 Parameters & Prior\\
\hline
Period $P$ & Jeffery's (1, 6000)\\
RV semi-amplitude $K$ & Mod-Jeffery's (1.01(1), 1000)\\
$\sqrt{e}\ {\rm sin}\ \omega$ & Uniform (-1, 1)\\
$\sqrt{e}\ {\rm cos}\ \omega$ & Uniform (-1, 1)\\
RV Jitter $\sigma$ & Mod-Jeffery's (1.01(1), 100)\\
RV Offset $\gamma$ & Uniform (-300, 300)\\
\hline
\end{tabular}
\end{table}

\texttt{RadVel} employs Markov Chain Monte Carlo (MCMC) technique to sample the posterior distribution of orbital elements, and adopts robust criteria to assess convergence. The best-fit orbital parameters and their associated uncertainties are derived from Maximum A Posteriori (MAP) fit.
Model selection (e.g. one-planet or two-planet model) is based on Bayesian Information Criteria (BIC; \citealt{Schwarz1978}) and reduced Chi-square $\chi_{\rm red}^{2}$. A criterion of $\Delta{\rm BIC}>5$ indicates two selected models show difference and the one with small BIC has the better goodness of fit.

It is known that the most significant limitation of RV method is the so-called $M_{\rm p}\,{\rm sin}\,i$ degeneracy, where $i$ is the orbital inclination. It means that RV method can only measure a minimum mass instead of true mass, simply because the observed quantities are radial velocities instead of true velocities. In order to remove the $M_{\rm p}\,{\rm sin}\,i$ discrepancy, we further utilize orbital fit code \texttt{orvara} \citep{Brandt2021b}, which jointly fits RV data and absolute astrometry from HGCA, to measure the true mass and inclination of planets. 

\texttt{orvara} was designed to fit Keplerian orbits to any combination of radial velocity, relative and/or absolute astrometry data. It uses the built-in package \texttt{htof} \citep{GMBrandt2021} to parse the Intermediate Astrometry Data (IAD) of Hipparcos, and then constructs covariance matrices to yield best-fit positions and proper motions of a star relative to the barycenter. Since the Gaia epoch astrometry or the along-scan residuals have not been released in the EDR3 and DR3, \texttt{htof} tentatively uses the synthetic data from Gaia Observation Forecast Tool \footnote{\url{https://gaia.esac.esa.int/gost/index.jsp}} (GOST) that contains the predicted observation time and scan angles to fit a 5-parameter astrometric model. The joint analysis of RV and Hipparcos-Gaia astrometry was extensively applied to determine the true mass of massive companions in recent years (e.g., \citealt{Liyt2021, Feng2022, Xiao2023}). The adopted priors and fitting strategies for \texttt{orvara} are detailed in \citet{Xiao2023}.

\begin{figure}[ht]
\plotone{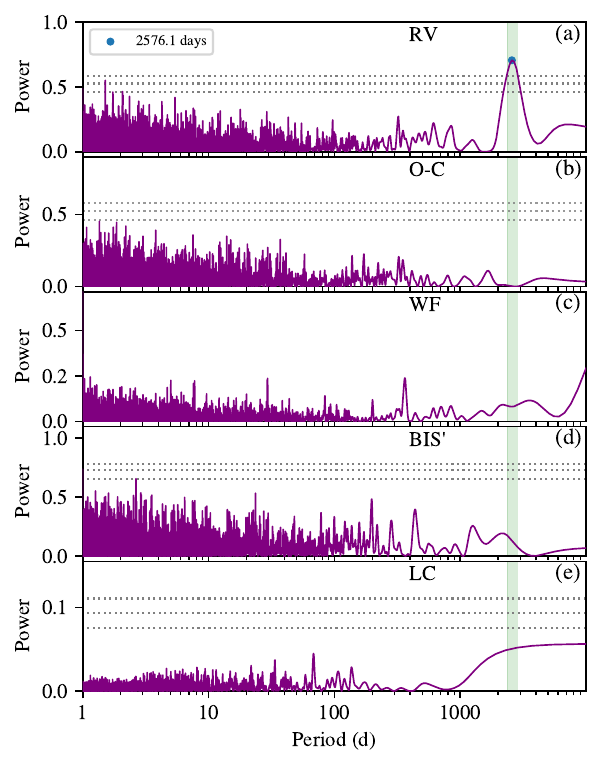}
\caption{Generalized Lomb-Scargle (GLS) periodograms for HD\,112570. (a) the GLS periodograms of the observed RVs. (b) the residuals to single Keplerian orbital fit (after subtracting the planet solution). (c) window function of sampling. (d) the mean-removed BIS (see Section \ref{sec:line}). (e) Hipparcos photometry. 
The horizontal grey lines, top to bottom, indicate the 0.001, 0.01, 0.1 False Alarm
Probability (FAP) levels, respectively. The green shaded area represents the period of planet signal. 
\label{fig:gls112570}}
\end{figure}

\begin{figure}[ht]
\plotone{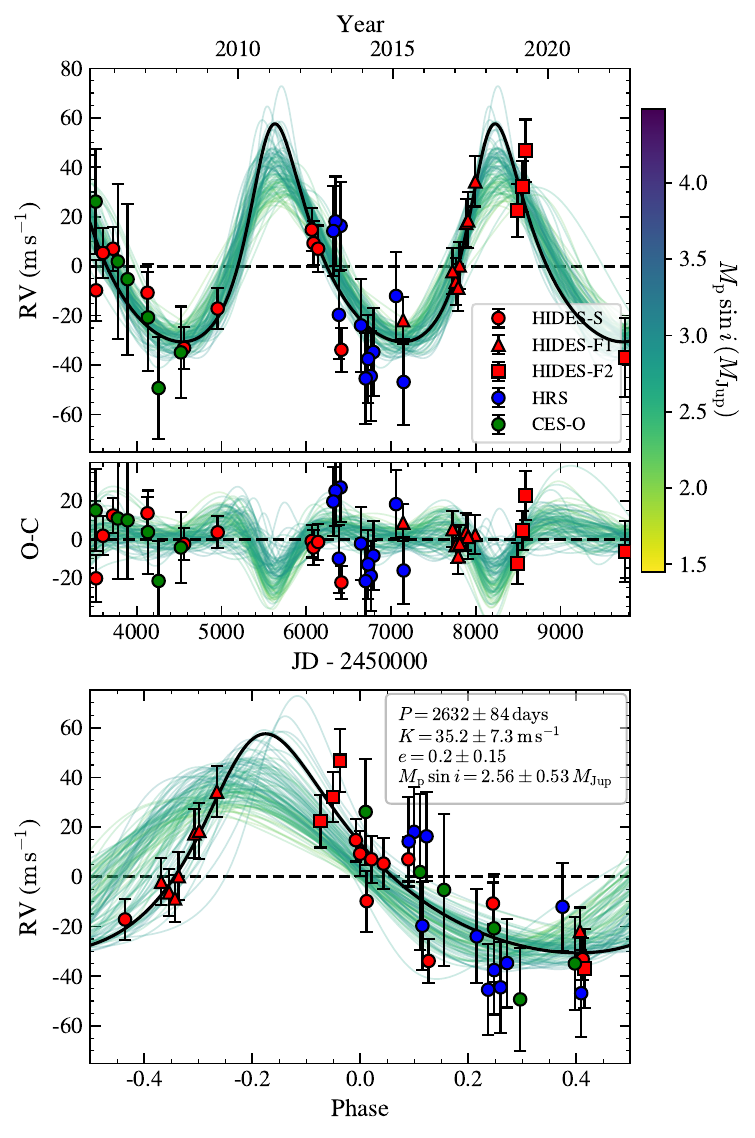}
\caption{\texttt{RadVel} results for HD\,112570.
Top panel: The best-fit single-planet Keplerian orbital model. The thick black line is the best-fit model, and the colored lines, color-coded by
the minimum mass of planet, indicate the possible orbital solution randomly drawn from the MCMC chain. Middle panel: the residuals to single Keplerian orbital fit (after subtracting the planet solution). Bottom panel: RVs phase-folded to the ephemeris of planet. The RV data from HIDES-S, -F1 and -F2 are shown in red circles, triangles and squares, respectively. The RV data from CES-O and HRS are shown in green and blue circles, respectively.
\label{fig:rv112570}}
\end{figure}

\subsection{HD\,112570} \label{subsec:hd112570}
Figure \ref{fig:gls112570} shows the GLS of HD\,112570 RVs. A moderately strong signal near 2576.1 days with FAP significantly lower than $0.1\%$ can be found. The best-fit single Keplerian orbit from \texttt{RadVel} yields a period of $P={2632}_{-78}^{+89}$ days, a RV semi-amplitude of $K={35.2}_{-6.4}^{+8.3}$ $\rm m\,s^{-1}$ and an eccentricity of $e={0.20}_{-0.14}^{+0.16}$. Given the stellar mass of $M_{\star}={1.15}_{-0.12}^{+0.12}\,M_{\odot}$ obtained by the global fit of \texttt{isochrones}, we derived a minimum mass of $M_{\rm p}\,{\rm sin}\,i={2.56}_{-0.48}^{+0.57}\,M_{\rm Jup}$ and a semi-major axis of $a={3.91}_{-0.16}^{+0.16}$ au for the Jovian planet HD\,112570\,b. The eccentricity is poorly constrained arising from the inadequate sampling near periastron.
Further RV monitoring will be required to fully constrain its orbit.
The Root Mean Square (RMS) of RV residuals is $\rm 13.16\,m\,s^{-1}$, slightly greater than the expected velocity amplitude ($\rm 9.02\pm1.28\,m\,s^{-1}$) of stellar oscillations derived from the scaling relation of \citet{Kjeldsen1995}. 
We found a $\Delta{\rm BIC}$ of 28 between the single planet model and the No-planet model, which indicates a distinct RV variation.
Furthermore, we did not find any correlation between RVs and brightness variation of the star based on Hipparcos photometry. The maximum rotational period of $P_{\rm rot}/{\rm sin}\,i\sim360\,{\rm days}$ estimated from the projected rotational velocity ($v\,{\rm sin}\,i=1.4\,{\rm km\,s^{-1}}$, \citealt{Glebocki2005}) and the radius of the host also excludes the possibility of false positive signal induced by stellar rotation.
In Figures \ref{fig:rv112570}, we plot the derived Keplerian orbits together with the RVs and associated uncertainties.

Besides, our joint analysis in combination of RV and HGCA astrometry also reveals a dynamical mass of $M_{\rm p}={3.42}_{-0.84}^{+1.4}\ M_{\rm Jup}$ and a prograde orbital inclination $i={54}_{-19}^{+23}\degr$ (or retrograde orbit ${126}_{-23}^{+19}\degr$). Other fitted or derived parameters from \texttt{orvara} are well consistent within $1\sigma$ with the values by \texttt{RadVel}. The detailed orbital parameters are listed in Table \ref{tab:orbital_par}. The plots of RV orbit and astrometric acceleration can be found in Figure \ref{fig:rv_pm_112570}, and the corner plot of orbital posteriors is shown in Figure \ref{fig:corner_112570}. All plots from \texttt{orvara} are presented in the Appendix \ref{sec:appendix_A}.

\subsection{HD\,154391}\label{subsec:hd154391}

\begin{figure}
\plotone{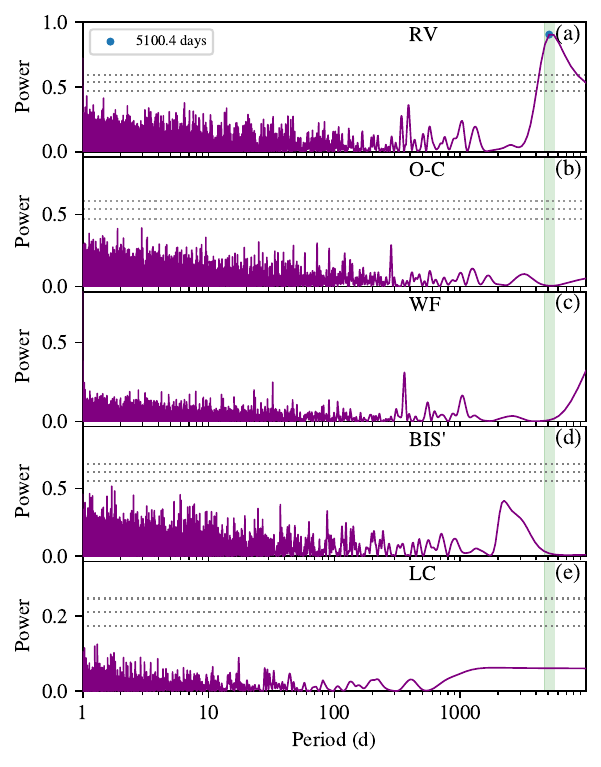}
\caption{Generalized Lomb-Scargle (GLS) periodograms for HD\,154391. The symbols are the same as Figure \ref{fig:gls112570}.
\label{fig:gls154391}}
\end{figure}

In Figure \ref{fig:gls154391}, We found a strong signal at 5100.4 days, indicating a periodic variation in HD\,154391 RVs. Therefore, a single Keplerian model was initially considered to fit the data. 

\begin{figure}
\plotone{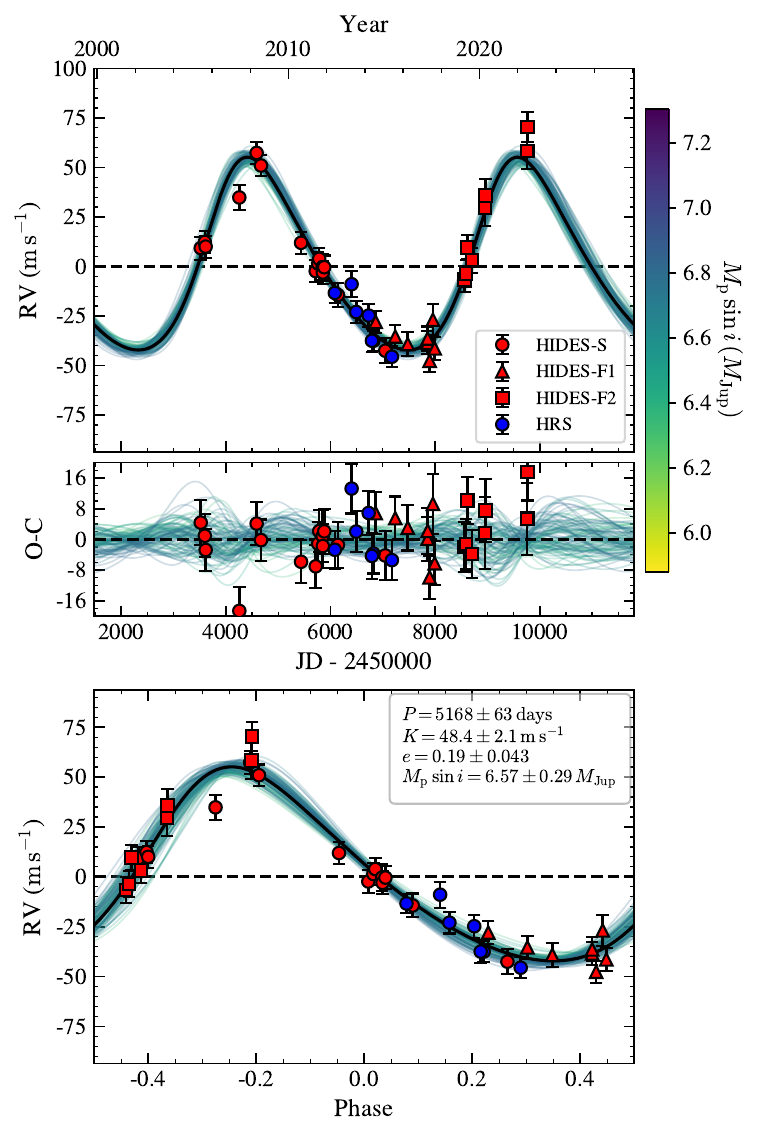}
\caption{\texttt{RadVel} results for HD\,154391. The symbols are the same as Figure \ref{fig:rv112570}.
\label{fig:rv154391}}
\end{figure}

Our best-fit solution from \texttt{RadVel} yields a period of $P={5168}_{-61}^{+65}$ days, a RV semi-amplitude of $K={48.4}_{-2.1}^{+2.2}$ $\rm m\,s^{-1}$ and an eccentricity of $e={0.19}_{-0.04}^{+0.04}$. Adopting a stellar mass of $M_{\star}={2.07}_{-0.03}^{+0.03}\,M_{\odot}$ derived by \texttt{isochrones}, we got a minimum mass of $M_{\rm p}\,{\rm sin}\,i={6.57}_{-0.29}^{+0.29}\,M_{\rm Jup}$ and a semi-major axis of $a={7.45}_{-0.07}^{+0.07}$ au for the super-Jupiter HD\,154391\,b. The orbital period is much greater than the maximum rotational period of $P_{\rm rot}/{\rm sin}\,i\sim430\,{\rm days}$. The RMS of RV residuals is $\rm 6.54\,m\,s^{-1}$, roughly corresponding to the expected velocity amplitude ($\rm 4.31\pm0.29\,m\,s^{-1}$) of stellar oscillations.

In addition, a slight linear trend which may be caused by an outer companion was also found. 
However, the difference of BIC value between no-trend and trend model is $\Delta{\rm BIC}=3.7$, suggesting 
the low possibility of the existence of extra companion. Furthermore, no evidence about its binarity can be identified in the Catalogue of the Components of Double and Multiple star (CCDM, \citealt{Dommanget2002}) or in the Washington Double Star Catalog (WDS, \citealt{Mason2001}).
We speculate that the trend should be attributed to the poorly observational quality in the recent exposures (e.g., bad weather condition). Consequently, we prefer the single-planet model without linear trend.
The best-fit orbits together with the RVs and uncertainties can be found in Figure \ref{fig:rv154391}.

Likewise, the RV data along with HGCA astrometry are also fit by \texttt{orvara} and the best-fit solution additionally reveals a mass of $M_{\rm p}={9.1}_{-1.9}^{+2.8}\ M_{\rm Jup}$ and an inclination of $i={47}_{-13}^{+21}\degr$ (or ${133}_{-21}^{+13}\degr$) for the planet.

\begin{table*}
\caption{Orbital Paremeters}
\label{tab:orbital_par}
\begin{tabular*}{\textwidth}{@{}@{\extracolsep{\fill}}lcccc@{}}
\hline\hline
Parameter & \multicolumn{2}{c}{HD\,112570\,b}  & \multicolumn{2}{c}{HD\,154391\,b}\\
& RV Only & RV + Astrometry & RV Only & RV + Astrometry\\
\hline

Period $P$ (days)        & ${2632}_{-78}^{+89}$        & ${2615}_{-77}^{+85}$               &  ${5168}_{-61}^{+65}$        & ${5163}_{-57}^{+60}$           \\
RV semi-amplitude $K$ ($\rm m\,s^{-1}$) & ${35.2}_{-6.4}^{+8.3}$            &${36.0}_{-6.1}^{+8.6}$     & ${48.4}_{-2.1}^{+2.2}$         & ${48.5}_{-2.0}^{+2.0}$        \\
$\sqrt{e}\ {\rm cos}\ \omega$ & ${0.28}_{-0.31}^{+0.22}$ & ${0.28}_{-0.33}^{+0.23}$& ${0.30}_{-0.07}^{+0.06}$ & ${0.28}_{-0.06}^{+0.06}$\\
$\sqrt{e}\ {\rm sin}\ \omega$ &${-0.16}_{-0.24}^{+0.31}$ & ${-0.14}_{-0.23}^{+0.31}$ & ${-0.31}_{-0.07}^{+0.10}$ &${-0.34}_{-0.07}^{+0.08}$\\

Eccentricity $e$ &  ${0.20}_{-0.14}^{+0.16}$          &${0.20}_{-0.14}^{+0.16}$                                 &  ${0.19}_{-0.04}^{+0.04}$        &${0.20}_{-0.04}^{+0.04}$         \\
Argument of periastron $\omega$ ($\degr$)&   ${334}_{-45}^{+75}$       &${301}_{-233}^{+40}$      & ${313}_{-11}^{+14}$         &${310}_{-10}^{+11}$       \\
Periastron time $T_{\rm p}$ (JD-2450000) & ${5618}_{-368}^{+943}$        &${5657}_{-262}^{+1156}$      &  ${9125}_{-154}^{+194}$        &${9075}_{-135}^{+161}$          \\
Semi-major axis $a$ (au) &  ${3.91}_{-0.16}^{+0.16}$      &${3.90}_{-0.16}^{+0.16}$                   & ${7.45}_{-0.07}^{+0.07}$         &${7.46}_{-0.07}^{+0.07}$            \\
$M_{\rm p}$\,sin\,$i$ ($M_{\rm Jup}$)&  ${2.56}_{-0.48}^{+0.57}$         &${2.62}_{-0.47}^{+0.58}$               & ${6.57}_{-0.29}^{+0.29}$         & ${6.59}_{-0.28}^{+0.28}$         \\
$M_{\rm p}$ ($M_{\rm Jup}$) &    —       &${3.42}_{-0.84}^{+1.4}$                    &    —      & ${9.1}_{-1.9}^{+2.8}$    \\
Inclination ($\degr$) &     —       &${54}_{-19}^{+23}$ (${126}_{-23}^{+19}$)      &    —      & ${47}_{-13}^{+21}$ (${133}_{-21}^{+13}$)     \\
Ascending node $\Omega$ ($\degr$) &  —    &${72}_{-49}^{+89}$                &     —     & ${106}_{-31}^{+32}$    \\
Semi-major axis (mas)&     —    &${38.8}_{-1.6}^{+1.6}$           &    —      & ${73.47}_{-0.65}^{+0.66}$        \\
Mass ratio &—&${0.00285}_{-0.00070}^{+0.0011}$&—&${0.00417}_{-0.00088}^{+0.0013}$\\
RV Jitter ${\sigma}_{\rm HIDES}$ ($\rm m\,s^{-1}$) & ${10.91}_{-2.58}^{+3.35}$         &${10.3}_{-2.4}^{+3.0}$     &  ${5.91}_{-1.19}^{+1.43}$        & ${5.6}_{-1.2}^{+1.4}$     \\
RV Jitter ${\sigma}_{\rm HRS}$ ($\rm m\,s^{-1}$) &  ${19.26}_{-7.66}^{+7.81}$           &${17.7}_{-4.1}^{+5.9}$     &   ${0.00}_{-0.00}^{+2.08}$       & ${0.42}_{-0.42}^{+5.3}$      \\
RV Jitter ${\sigma}_{\rm CES\text{-}O}$ ($\rm m\,s^{-1}$) &   ${0.001}_{-0.001}^{+2.62}$         &  ${0.012}_{-0.011}^{+1.7}$    &    —      &    —    \\
RV Offset ${\gamma}_{\rm HIDES}$ ($\rm m\,s^{-1}$) & ${-3.85}_{-4.00}^{+3.98}$         &${-4.4}_{-3.3}^{+2.9}$         &  ${-3.30}_{-1.34}^{+1.38}$        &  ${-3.35}_{-0.41}^{+0.39}$    \\
RV Offset ${\gamma}_{\rm HRS}$ ($\rm m\,s^{-1}$) & ${19.41}_{-7.66}^{+7.81}$           &${20.0}_{-4.1}^{+4.0}$       &   ${28.29}_{-3.53}^{+3.76}$       &  ${27.2}_{-2.3}^{+2.1}$     \\
RV Offset ${\gamma}_{\rm CES\text{-}O}$ ($\rm m\,s^{-1}$) &  ${139.48}_{-10.73}^{+10.71}$          &${139.7}_{-4.2}^{+4.7}$        &    —      &   —  \\
RMS ($\rm m\,s^{-1}$)&13.16&—&6.54&—\\
\hline
\end{tabular*}
\end{table*}

\section{Line profile and chromospheric activity} \label{sec:line}
Since the deformation of spectral line profile can result in wavelength shift and false positive planetary signal, we perform line profile analyses 
using the Bisector Inverse Span (BIS: \citealt{Dall2006}) as an indicator of line profile asymmetry. In a same manner with \citet{Takarada2018} and \citet{Teng2022}, the analysis method uses the iodine-free spectra in a wavelength range of 4000$-$5000 \AA\,to calculate the weighted cross-correlation function (CCF: \citealt{Pepe2002}) with a numerical mask of G-type giant star, which generated from \texttt{SPECTRUM} tool \citep{Gray1994} including about 800 absorption lines. The CCF profile represents the average profile of stellar absorption lines in a specific wavelength range. Then the BIS is defined as 
\begin{equation}
{\rm BIS} = v_{\rm top} - v_{\rm bot},
\end{equation}
where $v_{\rm top}$ denotes the average velocity of the bisectors' top region ($5\%-15\%$ from the continuum of CCF) and 
$v_{\rm bot}$ denotes the average velocity of the bisectors' bottom region ($85\%-95\%$ from the continuum of CCF). In addition, the different instrumental profile can lead to BIS offset, we therefore define the mean-removed BIS as 
\begin{equation}
{\rm BIS'} = {\rm BIS} - \overline{\rm BIS},
\end{equation}
to eliminate the difference within four instruments when we calculate Pearson's correlation coefficient of BIS and RVs. $\overline{\rm BIS}$ is the mean BIS of each instrument (HIDES and HRS).

In some cases, stellar chromospheric activities can masquerade as planetary signals, and we thus need to quantitatively evaluate the correlation between chromospheric activities and RV variations.
The flux of line core of Ca \uppercase\expandafter{\romannumeral2} HK lines is widely used to indicate the strength of chromospheric activities. Unfortunately, we only use H lines from HIDES-S spectra in this work due to the low signal-to-noise ratio of K lines. Additionally, both H and K lines are unavailable in HIDES-F1 and -F2 spectra due to severe aperture overlaps among $3700\sim4000$ \AA\,(see Section \ref{subsec:oao}).

Following \citet{Sato2013}, the Ca \uppercase\expandafter{\romannumeral2} H index is defined as 
\begin{equation}
{S_{\rm H}} = \frac{F_{\rm H}}{F_{\rm B}+F_{\rm R}},
\end{equation}
where $F_{\rm H}$ is the integrated flux of a 0.66 \AA\,wavelength region centered on the H line, while $F_{\rm B}$,
$F_{\rm R}$ denote the integrated fluxes
within a 1.1 \AA\,wavelength region centered on minus and plus 1.2 \AA\,from the H line, respectively. The uncertainties of ${S_{\rm H}}$ are estimated based on photo noise.

\begin{figure*}
\plotone{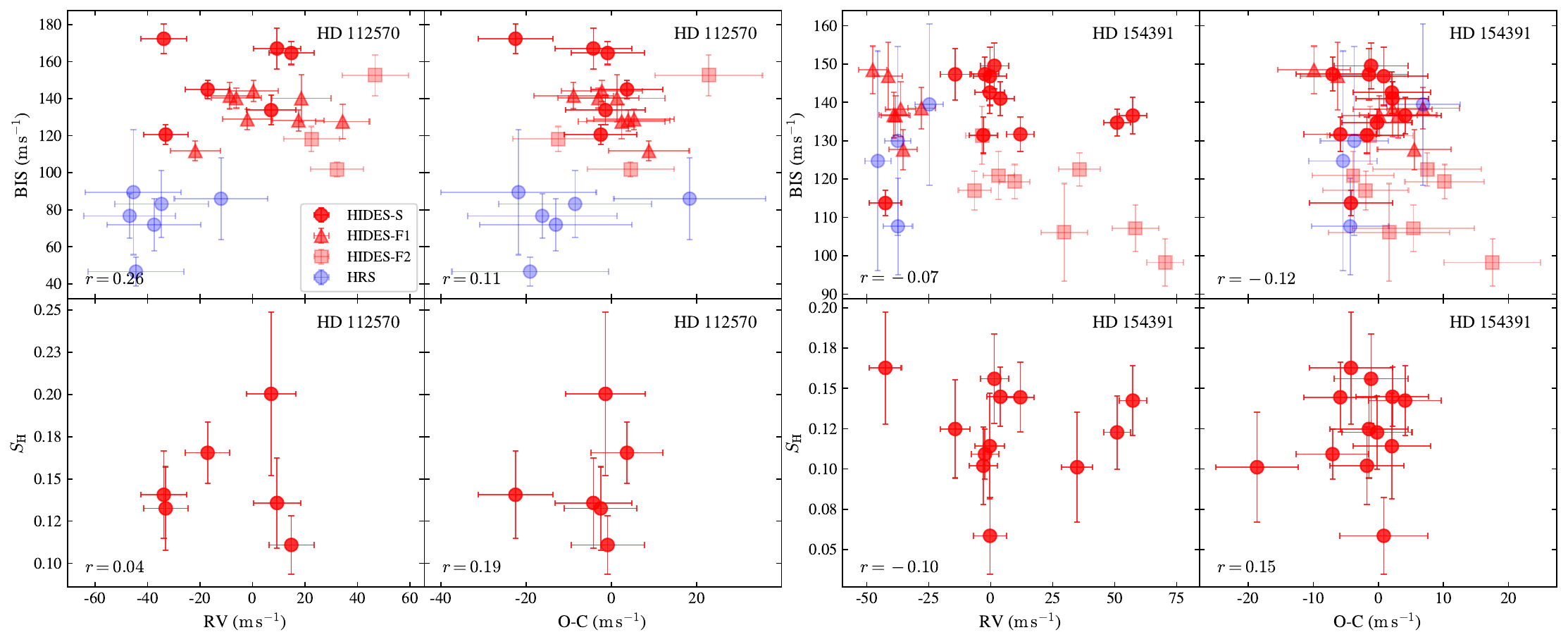}
\caption{BIS and ${S_{\rm H}}$ against RVs and 1-Keplerian fittings (O-C). HIDES-S data are shown in red circles, HIDES-F1 data are shown in red triangles and HIDES-F2 data are shown in red squares. HRS data from Xinglong are shown in blue circles. 
\label{fig:bis_shk}}
\end{figure*}

Figure \ref{fig:bis_shk} plots BIS and ${S_{\rm H}}$ against the observed RVs and RV residuals for two planetary systems. For HD\,112570, the large BIS offset between HIDES and HRS suggests that the line profile asymmetry is dominated by the instrumental profile, while the profile of HD\,154391 seems to be primarily attributed to stellar surface modulation. In Table \ref{Tab:pearson}, We further calculate Pearson's correlation coefficient $r$ of ${\rm BIS'}$ and ${S_{\rm H}}$ with RVs. All the values are within $-0.3\,\lesssim r \lesssim \,0.3$, and therefore show no correlation with RVs. In addition, the GLS periodogram of Hipparcos photometry of two stars don't exhibits any significant signal in corresponding planetary signals. Consequently, we can conclude that the regular RV variations of two stars are caused by orbital motion, rather than the deformation of spectral line profile and stellar chromospheric activity.

\begin{table}
\centering
\caption{Pearson's Correlation Coefficient}\label{Tab:pearson}
\begin{tabular}{lcccc}
\hline \hline
 Name & \multicolumn{2}{c}{HD\,112570} & \multicolumn{2}{c}{HD\,154391}\\
 & RVs &O-C &RVs&O-C\\
\hline
${\rm BIS'}$&0.22&0.16&0.08&0.08\\
${S_{\rm H}}$&0.04&0.18&$-0.11$&0.11\\
\hline
\end{tabular}
\end{table}

\section{Discussion} 
\label{sec:discus}
\subsection{Lithium Abundances}\label{subsec:lithium}
Lithium (Li) could be produced during nucleosynthesis in the early universe and was broadly used to study the mixing process of chemical compositions for stars. 
The standard stellar evolutionary theory predicts that the Li abundance of solar type stars at early RGB phase has been depleted from an initial meteoritic value of $A{\rm (Li)}\sim3.3\,{\rm dex}$ (e.g., \citealt{Grevesse1989}) to the current observation of $A{\rm (Li)}\sim1.5\,{\rm dex}$ \citep{Iben1967}, mainly attributing to the dredge-up process introduced by expanding convective envelope. 
As stars evolve near the RGB luminosity tip, Li surface abundance will be further diluted to $A{\rm (Li)}\sim0.5\,{\rm dex}$. However, a small fraction of red giant stars are known to be Li-rich stars with $A{\rm (Li)}>1.7\,{\rm dex}$ (e.g., BD\,+48\,740, $A{\rm (Li)}=2.3\,{\rm dex}$, \citealt{Adamow2012}), and thus several studies about Li enrichment mechanisms, e.g., planetary engulfment events \citep{Kunitomo2011,Aguilera-Gomez2016, Soares-Furtado2021, Behmard2023}, have arisen. 

Using spectral synthesis method, \citet{Liu2014} measured $A{\rm (Li)}=0.71\,{\rm dex}$ for the RGB star HD\,112570, and an upper limit of $-0.07\,{\rm dex}$ for the RC star HD\,154391, respectively. Both giant stars show severe Li depletion. The former appears to agree with expectations from stellar evolutionary theory, while the latter implies Li depletion may have already take place in the MS phase or may be more efficient for giants undergoing helium flash.
Several depletion mechanisms driving the Li dip are proposed, such as the material mixing induced by overshooting between interior and convective envelope, the mixing induced by rotation, and exterior celestial body (e.g., \citealt{Brun1999,Xiong2009,Fu2015}). 
In addition, some studies suggested that Li abundance is linked with the presence of planets (e.g., \citealt{Israelian2004,Israelian2009,Takeda2005,Chen2006,Liu2014}). Other studies found no such connection (e.g., \citealt{Baumann2010}). For example, \citet{Liu2014} found Li abundance is easy to deplete for giant stars harboring planets, while \citet{Adamow2018} reported a relatively high frequency of planets around Li-rich giant stars. Our two planetary systems seem to support the finding of \citet{Liu2014}.  
Further rigorous analysis is beyond the scope of this paper.

\subsection{Planetary Formation}\label{formation}

\begin{figure*}[!ht]
\plotone{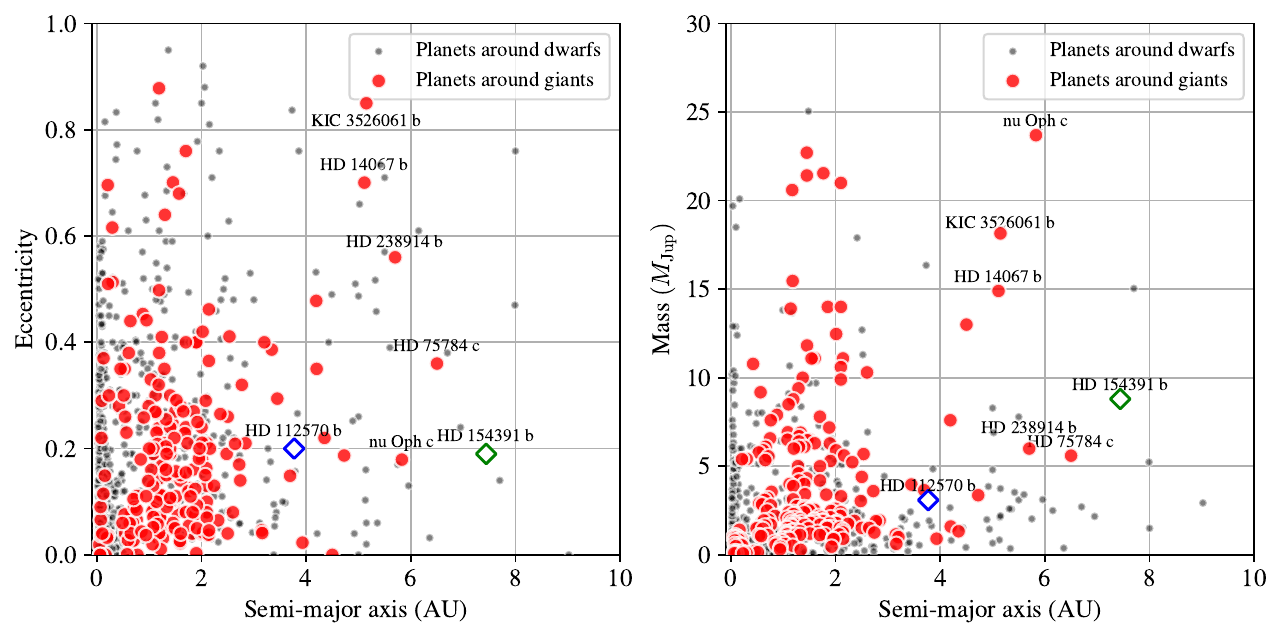}
\caption{Left panel: Eccentricity versus semi-major axis ($a<10\,$au) for known giant planets ($M_{\rm p}\,{\rm sin}\,i>0.1\,M_{\rm Jup}$) around dwarf and giant stars. Right panel: Planetary (minimum) mass versus semi-major axis. HD\,112570\,b and HD\,154391\,b are marked by blue and green diamond. The grey dots and red circles correspond to dwarf and giant hosts, respectively. We also show five substellar companions with orbits at the most wide separation. HD\,154391\,b has one of the longest orbital period among known planets orbiting evolved stars. All the data points are compiled from the \href{http://exoplanet.eu/catalog/}{exoplanet.eu} \citep{Schneider2011} and NASA exoplanet archive \citep{Akeson2013}. Data acquisitions on April 2, 2023.
\label{fig:mass_ecc_period}}
\end{figure*}

\begin{figure}[ht!]
\plotone{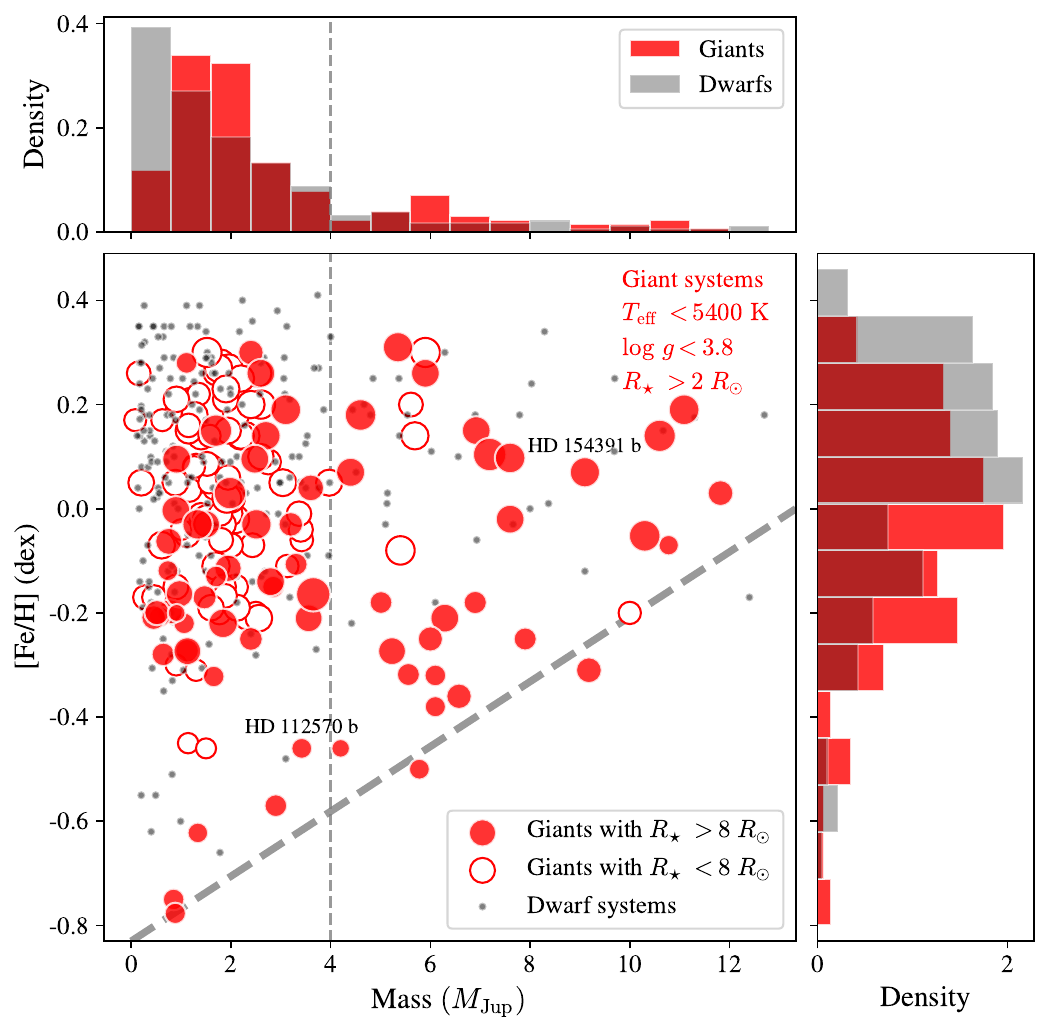}
\caption{Stellar metallicity with respect to planetary (minimum) mass ($M_{\rm p}\,{\rm sin}\,i<13\,M_{\rm Jup}$). The red circles (both filled and empty circles) correspond to planet-hosting giant stars, and grey dots correspond to dwarf stars. Giant stars with stellar radius below and above $8\,R_{\sun}$ (see Figure \ref{fig:r_a}) are indicated in red empty and filled circles, respectively, and the size of them is proportional to stellar mass.
It is worth noting that giants with radius $>21\,R_{\sun}$ are excluded in this figure, because the planets round them might be masqueraded by stellar intrinsic variability (see next section). The top and right panel show histograms of planetary mass and stellar metallicity, respectively. An evident transition at $\sim 4\,M_{\rm Jup}$ can be found within the histogram of giants in top panel.
The thick dashed line in grey denotes a possible limit for giant stars in metallicity-mass plane.
\label{fig:feh_mass}}
\end{figure}

\begin{table*}[!t]
\caption{The result of the KS tests for giant stars ($R_{\star}<21\,R_{\sun}$)}
\label{tab:ks_test}
\begin{tabular*}{\textwidth}{@{}@{\extracolsep{\fill}}lccccc@{}}
\hline\hline
Parameter & \multicolumn{2}{c}{$M_{\rm p}<4\,M_{\rm Jup}$ (126)}  & \multicolumn{2}{c}{$M_{\rm p}>4\,M_{\rm Jup}$ (32)} & KS $p$-Value\\
&Mean&STD&Mean&STD&\\
\hline
Stellar Mass $M_{\star}$ ($M_{\sun}$)&1.49&0.33&1.67&0.47&$\mathbf{0.004}$\\
Stellar Radius $R_{\star}$ ($R_{\sun}$)&7.09&3.51&11.92&4.44&$\mathbf{5.65\times10^{-7}}$\\
Metallicity $\rm [Fe/H]$&$-0.03$&0.22&$-0.07$&0.23&0.160\\
Eccentricity $e$&0.17&0.15&0.22&0.19&0.451\\
Period $P$ (days)&661.78&567.96&804.00&1083.31&0.466\\
\hline
\end{tabular*}
\tablecomments{The $p$-values smaller than 0.05 are highlighted with boldface. The count of each sub-sample is presented in parenthesis of the first row.}
\end{table*}

Both the two planets have moderate eccentricity and long orbital period, especially for HD\,154391\,b which has one of the longest period ($>5000\,{\rm days}$) among those ever found around evolved stars (Figure \ref{fig:mass_ecc_period}). 
Unlike $\iota$ Draconis c ($M_{\rm p}={17.0}^{+13}_{-5.4}\,M_{\rm Jup}$, $a={19.4}^{+10}_{-7.7}\,{\rm au}$, $P={68}^{+60}_{-36}\,{\rm yr}$), a substellar companion found orbiting a K giant star with the longest period but loosely being constrained by insufficient RV sampling and HGCA astrometry \citep{Hill2021}, our orbital solution for HD\,154391\,b is quite well-characterised by the complete orbital phase coverage. 
Considering the low occurrence rate in wide orbit which is far beyond the snow line, those giant planets may rise the interest about their formation and dynamical history.

In the past three decades, two main mechanisms were developed to explain the formation of giant planets. 
One is the core accretion model (e.g., \citealt{Pollack1996, Ida2004,Santos2004}), a bottom-up process that begins with the formation of a massive core of $5\sim20\,M_{\oplus}$, followed by the rapid accumulation of a gaseous envelope from the protoplanetary gas disk. The other one is disk instability model \citep{Boss1997}, a top-down process that directly forms through the gravitational collapse of a protoplanetary disk. The core accretion model can well explain the well-known correlation between giant planet occurrence and stellar metallicity \citep{Fischer2005}, while disk instability model may lead to the formation of more massive companions in metal-poor conditions \citep{Meru2010}. 

\subsubsection{Planetary Mass}
Recently, some works pointed out that the different masses of giant planets may indicate the different formation mechanisms. \citet{Santos2017} explored the properties of the minimum mass (or mass) and metallicity distribution of giant planets discovered through RV and transit methods. Using the homogeneously derived host metallicities in SWEET-Cat \citep{Santos2013}, they found two distinct populations separated by $\sim4\,M_{\rm Jup}$, and thus proposed that giant planets with $M_{\rm p}\,{\rm sin}\,i< 4\ M_{\rm Jup}$ are primarily formed by core-accretion mechanism, while giant planets with $M_{\rm p}\,{\rm sin}\,i> 4\ M_{\rm Jup}$ are formed by disk instability. 
\citet{Teng2022} reported a similar mass gap for giant planets around intermediate-mass stars, while \citet{Adibekyan2019} found no evidence about the existence of a breakpoint at $4\ M_{\rm Jup}$, and argued that planets with same (high) mass can be formed through different channels depending on the specific disk conditions (e.g., disk lifetime and mass). 

Figure \ref{fig:feh_mass} shows host metallicity with respect to (minimum) mass of currently known giant planets discovered by RV method. Evolved stars defined in this paper are restricted to $T_{\rm eff}<5400\,{\rm K}$, ${\rm log}\,g<3.8$ and $R_{\star}>2\,R_{\sun}$. 
All the data points are compiled from the \href{http://exoplanet.eu/catalog/}{exoplanet.eu} \citep{Schneider2011} and NASA exoplanet archive \citep{Akeson2013}. Giant stars with radius $>21\,R_{\sun}$ are excluded in this figure, since their intrinsic jitter might mimic planetary signals (see next section).
A clear gap at $4\ M_{\rm Jup}$ can be seen in top panel (red histogram), even if we assume random orientations of inclination, which appears to agree with \citet{Santos2017} and \citet{Teng2022}. Our two findings reside in the opposite sides of this gap, implying core accretion might be responsible for the formation of HD\,112570\,b (${3.42}_{-0.84}^{+1.4}\ M_{\rm Jup}$) and disk instability for HD\,154391\,b (${9.1}_{-1.9}^{+2.8}\ M_{\rm Jup}$).
Interestingly, this gap only emerges in the mass distribution of planets around giants, while weakening for dwarf systems. 
Besides, a rough limit (thick dashed line) can also be identified, suggesting the lower metallicity of stars, the less massive planets they may host, i.e., very massive giant planets cannot form at low metallicities. 
Despite the low average metal abundance of giant hosts, this limitation appears to be consistent with the general prediction of core accretion paradigm (e.g., \citealt{Fischer2005, Mordasini2012}). 
We also note that some of the hosts with ${\rm [Fe/H]}<-0.2$ have enhanced $\alpha$-elements (e.g., Mg, Si, Ca, Ti), making them not so much metal-poor but iron-poor.
Finally, it is worth noting that the gap at $4\ M_{\rm Jup}$ will widen and the limit envelope will weaken with the addition of larger giant stars ($>21\,R_{\sun}$). 

To further explore whether the gap can reveal two different formation channels for planets orbiting giants, we perform Kolmogorov-Smirnov (KS) tests\footnote{Using the python \texttt{scipy.stats.ks\_2samp} library} for these two populations in other parameter spaces. 
Table \ref{tab:ks_test} lists the result of this analysis, and shows a probability of $16\%$ for the two subsamples drawing from the same underlying metallicity distribution. 
It suggests that the mass gap at $4\ M_{\rm Jup}$ can not reveal two distinct populations in metallicity \citep{Adibekyan2013}, although the average metallicity of giant stars with more massive planets ($M_{\rm p}>4\,M_{\rm Jup}$) is slightly lower.
However, KS tests also suggest that the differences of stellar mass and radius between the two populations are statistically significant ($p<0.05$), i.e., more massive planets orbit around, on average, heavier and larger giant stars. As can be seen from Figure \ref{fig:feh_mass}, planets with mass larger than $4\ M_{\rm Jup}$ have almost be found orbiting larger stars ($>8\,R_{\sun}$). This tendency should attribute to inadequate sampling or detection biases rather than real physics, e.g., larger stars tend to exhibit larger intrinsic jitter, and thus lead to the discovery of more massive planets who can introduce relatively larger RV amplitude. In addition, assuming the disk mass is scaled roughly linearly with stellar mass, more massive hosts might harbour more massive planets. 
Therefore, the observational gap probably suffers detection biases or small number statistics, at least we cannot verify the existence of two distinct planet populations according to current sample. A larger and non-biased database is required to confirm this hypothesis. 

\citet{Schlaufman2018} also studied the relation between host metallicity and planetary mass characterized by both RV and transit methods. He found a mass limit of $M_{\rm p}\sim10\ M_{\rm Jup}$ within the homogeneous sample, and suggested that planets formed by core accretion have a maximum mass of no more than 10 $M_{\rm Jup}$, i.e., planets do not prefer orbiting metal-rich hosts above this limit. However, some studies speculated that core accretion may even work on low-mass brown dwarf region (e.g., \citealt{Maldonado2017,Xiao2023}), and planet population syntheses (e.g., \citealt{Mordasini2012, Emsenhuber2021}) also predicted that more massive companions can be formed via core accretion at more wider separation than Jupiter-like planets. According to above discussions, the masses of HD\,112570\,b and HD\,154391\,b seem to place them to the systems likely formed via core accretion.

\begin{figure*}[ht]
\plotone{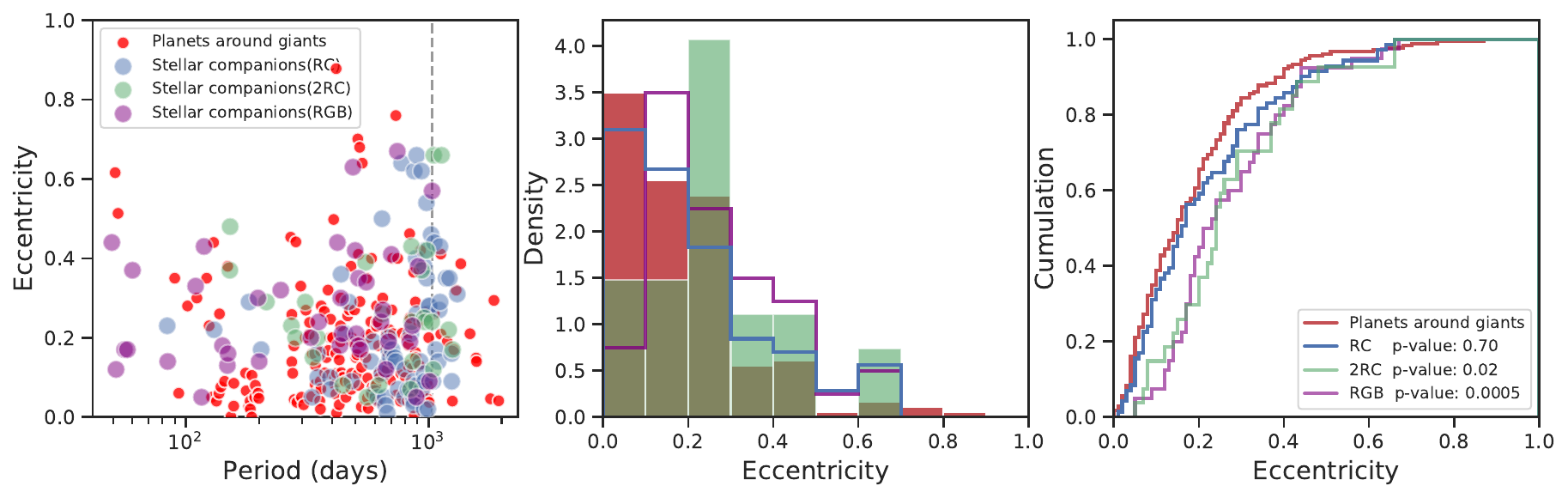}
\caption{Comparison of the evolved stellar binaries with giant planets orbiting giant star. Left panel: Eccentricity-period distribution. The light blue, green and purple circles denote the evolved systems with the primary star settling on the RC, 2RC and the RGB phase, respectively. The vertical dashed line denotes the 1034 day baseline of Gaia DR3. The period of planets is restricted to $50\sim2000\,$ day to match the period distribution of stellar binaries. Middle panel: Histogram of eccentricity distribution. Right panel: Cumulative distribution. The legend shows the $p$-values between planets and stellar binaries. The eccentricity distribution of planets orbiting giant stars is similar with RC. The data of evolved stellar binaries are compiled from Appendix B of \citet{Beck2023arXiv}.
\label{fig:per_ecc_beck}}
\end{figure*}

\subsubsection{Eccentricity}
On the other hand, the consideration of extra parameter spaces, such as eccentricity and orbital period, may also provide crucial clues for planetary formation.  
After the formation of planets within the disk, their eccentricity may be excited by several possible mechanisms: planet-planet scattering (e.g., \citealt{Ford2008}), secular Kozai-Lidov perturbations induced by outer massive companions (e.g., \citealt{Lidov1962,Kozai1962,Naoz2016}), and planet-disk interactions (e.g., \citealt{Goldreich2003}). 
Different distribution of eccentricity may indicates different formation scenarios.  According to the population-level eccentricity analysis of 27 long-period and directly imaged substellars between 5 and 100\,au,  \citet{Bowler2020} found companions with $M_{\rm p}<15\,M_{\rm Jup}$ or $M_{\star}/M_{\rm p}<0.01$ show lower orbital eccentricity, while brown dwarfs with $M_{\rm p}=15\sim75\,M_{\rm Jup}$ or $M_{\star}/M_{\rm p}=0.01\sim0.2$ exhibit higher eccentricity. They explained it as the evidence for imaged planets formed via core accretion, and for brown dwarfs formed via molecular cloud fragmentation.
The mean eccentricity of their single long-period planets is $\bar{e}=0.23$, comparable to the value of 0.20 for HD\,112570\,b and 0.19 for HD\,154391\,b, respectively. This may further supports the core accretion scenario for our two planets.
In addition, we also make a comparison of planet population around giant stars with evolved stellar binaries.
\citet{Beck2023arXiv} made a comprehensive study of stellar companions of giant stars. They found the RGB and secondary clump (2RC, $M_{\star}\gtrsim2\,M_{\odot}$) stars retain a relatively flat eccentricity distribution, while the RC ($M_{\star}\lesssim2\,M_{\odot}$) stars have highest occurrence rate below $e\lesssim0.2$ likely due to the accumulated tidal effect along the stellar evolution. At the tip of the RGB, a star with $M_{\star}\lesssim2\,M_{\odot}$ might expand to a maximum radius of $\sim175\,R_{\odot}$ that is expected to promote orbital circularization.
As shown in Figure \ref{fig:per_ecc_beck}, planets around giant stars tend to have lower eccentricity, and show similar distribution with the RC stars ($p=0.70$). 
It seems that some planets might have undergone tidal interaction to damp eccentricities like RC stars. 
However, given that there are several uncertain quantities (e.g., evolutionary state, age and tidal quality factor) for planet-hosting giant stars, it is hard to distinguish which fractions of planets are the product of tidal circularization, and which fractions are sculpted by other mechanisms or maintained primordial eccentricity.

\subsubsection{Possible Formation Pathway}
One may wonder if core accretion is suitable for HD\,154391\,b when considering its remarkably wide separation and the lifetime of the disk. We note the period dependence of giant planet occurrence revealed that the peak ($\sim6\,\%$) appears at $\sim800\,{\rm days}$, and dramatically drops to below $1\,\%$ for period above $\sim5000\,{\rm days}$ \citep{Wolthoff2022}. 
Although planets are rare in such wide orbit, \citet{Wagner2019} suggested that giant planets, similar to the close-in giant planets, can also form via core accretion at large orbital separation ($a\gtrsim8\,{\rm au}$). Moreover, the stellar mass $M_{\star}=2.07\,M_{\odot}$ of HD\,154391 may imply a massive and dense disk that supports the formation of a massive planet more easily (e.g., \citealt{Mordasini2012, Wolthoff2022}). 
Direct imaging surveys for planet around massive BA stars have also revealed that giant planets are commonly found around higher mass stars (e.g., \citealt{Nielsen2019, Vigan2021}).
However, it is undeniable that the short lifetime of massive disks may affect the formation efficiency of giant planets formed via core accretion, while disk instability without that. Disk instability is thought to operate far away from the central star ($>10\,{\rm au}$), that allows for more efficient cooling and collapse (e.g., \citealt{Rice2022, Meru2010}). Although the population synthesis model of \citet{Forgan2018} shows that it is possible for some companions to undergo inward migration \citep{Baruteau2011} and tidal disruption \citep{Nayakshin2017} to decrease their mass on a much closer-in orbit, this occurrence is extremely rare and the resulting orbits tend to exhibit high eccentricity \citep{Rice2022, Matsukoba2023}. Those companions formed via disk instability tend to remain relatively massive and be on relatively wide orbits ($>10\,{\rm au}$). Consequently, it seems more likely that HD\,154391\,b ($7.45\,{\rm au}$) was formed through core accretion. 

As for HD\,112570\,b, its host has an extremely low metallicity (${\rm [Fe/H]=-0.46}$). One may doubt about whether it can be formed via core accretion in such a metal-poor condition \citep{Ida2004}. Previous work of \citet{Liu2010} has recognized HD\,112570 as a thin disk star and found no any indications of $\alpha$-elements enhancement ($[\alpha/{\rm Fe}]=0.06$). Those metal-poor stars were thought to migrate from the outer disk of the galaxy, where the low abundance of $\alpha$-elements content may hinder the rocky planetesimals/planets from growth within the circumstellar disk \citep{Haywood2009,Adibekyan2012}. However, just as mentioned above, it is hard for disk instability to form a relatively closer planet ($<10\,{\rm au}$). We note the synthetic planets based on core accretion model of \citet{Mordasini2012} suggested those massive planets orbiting extremely metal-poor stars (see Figure \ref{fig:feh_mass}) might indicate the boundary where core accretion can be in operation. Therefore, difficulties remain with both formation models for HD\,112570\,b at this time. Further development of the formation model is awaited.

\subsection{Stellar Radius$-$Planetary Semi-major Axis Diagram}\label{radius_axis}

\begin{figure}[ht!]
\plotone{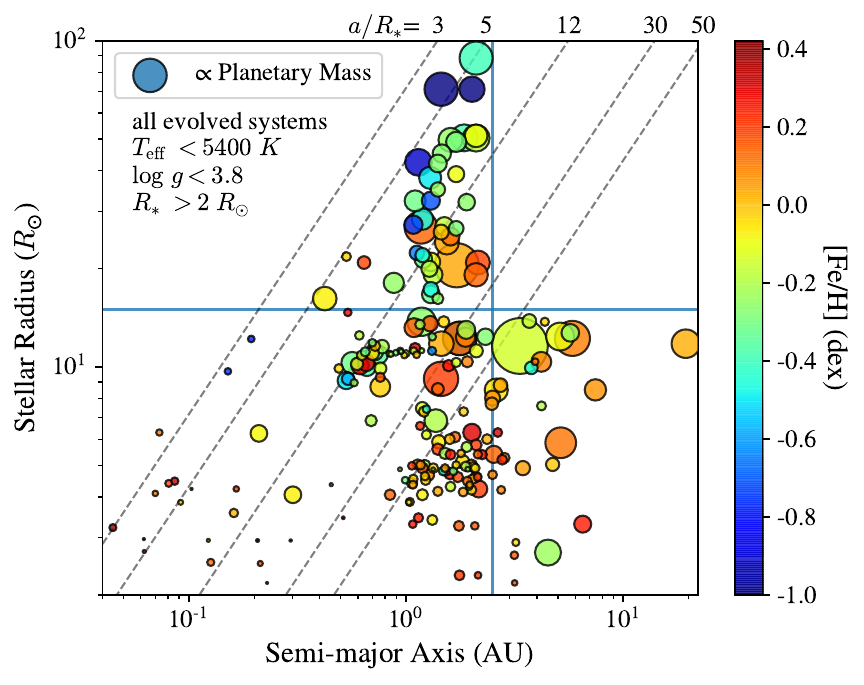}
\caption{Stellar radius with respect to planetary semi-major axis. The circles represent currently known evolved systems. The size and color of the circles are scaled by planetary (minimum) mass and stellar metallicity, respectively. The oblique dashed lines indicate the ratios of semi-major axis to stellar radius. The horizontal and vertical cyan lines denote $R_{\star}=15\,R_{\sun}$ and $a=2.5\,$au, respectively.
\label{fig:r_a}}
\end{figure}

In stellar radius and planetary period plane for planets around giants, \citet{Dollinger2021} found that there is a clear deficit of short-period ($\lesssim300\,{\rm days}$) and long-period ($\gtrsim800\,{\rm days}$) planets around giants with radii greater than $21R_{\odot}$, i.e., planets prefer similar periods above a certain radius. As shown in Figure \ref{fig:r_a}, we demonstrate this finding in radius$-$semi-major axis plane.
They stated that those planets are probably false positives masqueraded by a known/unknown phenomenon, such as rotational modulation or a kind of stellar oscillations. 
In addition, they also found the apparent accumulation can only happen for unbiased sample, i.e., without $B-V<1.2$ cutoff, as biased sample has almost excluded larger giant stars. Although this cutoff for most surveys of planets around giants can minimize intrinsic stellar jitter, it may give rise to statistically biased results (e.g., planet-metallicity correlation for giant stars).

Assuming those planets orbiting larger giant stars are indeed bona fide planets, we find the lack of short-period planets may be naturally explained by star-planet interactions (e.g., engulfment). 
As shown in Figure \ref{fig:r_a}, there is a rough limit of $a/R_{\star}=3\sim 5$ in the radius-axis plane, which may imply that it is hard for close-in planets to survive long. In addition, we find the mean mass of planets around larger giant stars is nearly twice as much as those planets orbiting smaller giant stars.
This may suggest a growing mechanism for planets during the giant phase of the larger stars. But we cannot exclude the possibility that the RV amplitude introduced by smaller planets is far below the stellar intrinsic jitter (e.g., oscillations, \citealt{Kjeldsen1995}). 
\citet{Jones2014} put forward some possibilities to interpret the high fraction of massive planets around post-MS stars. They thought that planets can directly accrete material from stellar envelope or stellar wind during the RGB phase. This may be more efficient for planets around larger giant stars due to their hosts' larger mass loss. As for the lack of long-period planets, the limit of stellar radius seems to extend to $R_{\star}=15\,R_{\odot}$. Likewise, the low RV amplitude and the large stellar jitter may be responsible for this deficit. For a specific star ($M_{\star}=1.5\,M_{\sun}$, $R_{\star}=20\,R_{\sun}$, $T_{\rm eff}=4300\,{\rm K}$), the expected RV variation solely induced by stellar oscillations is $\rm \sim40\,m\,s^{-1}$, roughly corresponding to a $3\,M_{\rm Jup}$ planet in 2.5\,au orbit \citep{Kjeldsen1995}. Therefore, it is more difficult to detect wide-orbit planets around large stars. 
However, we are still not clear what mechanism can cause the pile of planets around larger giant stars? One possibility may relate to the halt of inward migration for giant planets when the circumstellar disks quickly dissipated. The theoretical analysis is beyond the scope of this paper.

What if those planets are not real? Is there a new phenomenon hidden in stars that is beyond our current understanding? 42\,Dra, a K giant with radius of $22\,R_{\sun}$, was reported hosting a $3.9\,M_{\rm Jup}$ planet in a 479.1 days orbit \citep{Dollinger2009}. However, the RV amplitude derived from follow-up RV measurements unexpectedly decreased by a factor of $\sim4$, challenging the existence of the reported planet \citep{Dollinger2021}. Similar behavior has been happened for $\gamma\,{\rm Dra}$, who initially shows regular RV variations with a period of 702 days from a decade of monitoring. Afterward, its RV variations suddenly disappeared and finally appeared again but with a phase shift \citep{Hatzes2018}. Oscillatory convection modes and rotational modulation have been suggested to explain the observed accumulation in several hundred days. \citep{Saio2015, Dollinger2021}. These cases may present a hint of unknown stellar variability. To answer above questions, it is apparent that more knowledges about the activities in large stars and further continuous RV monitoring are required. 

Additionally, apart from RV method, other techniques, such as astrometry, may be superior complements to uncover the realities of those doubtful planets. Recent Gaia astrometry has been used to reveal the real natures of giant planets. We thus make an inspection of the Renormalised Unit Weight Error (RUWE) for planet-hosting stars ($>15\,R_{\sun}$). In practice, it is widely accepted that a value of $\rm{RUWE}<1.4$ indicates a good astrometry solution, while $\rm{RUWE}>1.4$ implies the binarity or multiplicity of a star~\citep{Lindegren2018, GaiaCollaboration2022}.  
Six giant stars, i.e., $\beta\,{\rm Cnc}$ (2.95), $\tau\,{\rm Gem}$ (2.35), HD\,113996 (1.50), 42\,Dra (1.83), HD\,66141 (2.21) and 4\,UMa (1.79), have $\rm{RUWE}>1.4$. Their anomalies in RUWE may be attributed to the perturbation of unseen companions, which seems to match the existence of those reported planets. However, some of them are found hosting wide stellar companions, and we thus can't confirm that the anomalies in RUWE are indeed caused by their orbiting planets. We expect the epoch astrometry from Gaia DR4 can enable us to clarify those issues better.

\section{Summary} 
\label{sec:summary}
In this paper, we report the discoveries of two long-period giant planets, HD\,112570\,b and HD\,154391\,b, according to the precise RV measurements from Xinglong and OAO observatories. 
HD\,154391\,b has one of
the longest orbital periods (${5168}_{-61}^{+65}\,{\rm days}$) among those ever found around giant stars. Besides, We combined RV and Hipparcos-Gaia astrometry to derive their inclinations and masses which can constrain their real natures. The masses are found to be $M_{\rm p}={3.42}_{-0.84}^{+1.4}\ M_{\rm Jup}$ for HD\,112570\,b and $M_{\rm p}={9.1}_{-1.9}^{+2.8}\ M_{\rm Jup}$ for HD\,154391\,b, respectively, well residing in the planetary region.

Li abundance are thought to be linked with the chemical mixing process of stars, and the existence of planets. Both hosts, HD\,112570 and HD\,154391, are found having been suffered severe Li depletion, suggesting this depletion may have already take place in the MS phase. Besides, these two Li-depleted planetary systems seems to support the point that Li abundance is easy to deplete for giants harboring planets (e.g., \citealt{Liu2014}). Population-level analysis is recommended to corroborate this hypothesis.

Considering the low occurrence rate of planets in wide orbit,
HD\,154391\,b may rise the interest about its formation and dynamical history. The previously reported mass gap at $4\,M_{\rm Jup}$ is also evident for giant star population, but it seems that giant planets with mass bellow and above this limit show identical distribution in metallicity, implying their similar formation channel.
Based on the large sample studies from previous works (e.g., \citealt{Bowler2020, Wolthoff2022}), it seems more likely that core accretion scenario should be responsible for the formation of HD\,154391\,b. 
As for HD\,112570\,b, it grows in a metal-poor condition without apparent $\alpha$-elements enhancement, we believe it is still difficult for both models to interpret the formation of the planet.
In addition, both planets exhibit moderate eccentricity, implying that they might have not experienced active dynamics (e.g., planet-planet scatting) when compared with transiting evolved systems. 

Planets orbiting giants known today might have been contaminated by fake positives, owing to some known/unknown phenomena of stars mimicking planet signals. It might give rise to overestimated planetary occurrence rate for evolved systems, if researchers don't pay high cautions to their observed RV variations. 
The abnormal accumulation near 2 au for planets around large giants seems to provide a kind of hint for such intrinsic stellar variability \citep{Dollinger2021}. Unfortunately, with the limitation of the understanding for highly evolved stars, it is still difficult for us to reach a reality.  


\section{Acknowledgments}
We thank the anonymous referee for providing great suggestions to improve the paper.
We would like to express our appreciation to Shigeru Ida and for his valuable comments and suggestions on our manuscript. We thank Daniel Huber for developing the code in \citet{Hon2017, Hon2018} with Marc Hon. We also express our special thanks to Yoichi Takeda for the support in stellar property analysis.
This research is supported by the National Key R\&D Program of China No. 2019YFA0405102 and 2019YFA0405502. 
This research is also supported by the National Natural Science Foundation of China (NSFC) under Grant No. 12073044, 11988101, 12261141689, and U2031144. 
We acknowledge the support of the staff of the  Xinglong 2.16m telescope. This work was partially supported by the  Open  Project  Program  of  the  Key  Laboratory  of  Optical  Astronomy,  National Astronomical Observatories, Chinese Academy of Sciences.

The Okayama 188cm telescope is operated by a consortium led by Exoplanet Observation Research Center, Tokyo Institute of Technology (Tokyo Tech), under the framework of tripartite cooperation among Asakuchi-city, NAOJ, and Tokyo Tech from 2018.  H.Y.T appreciate the support by the EACOA/EAO Fellowship Program under the umbrella of the East Asia Core Observatories Association. B.S. was partially supported by MEXT's program ``Promotion of Environmental Improvement for Independence of Young Researchers'' under the Special Coordination Funds for Promoting Science and Technology, and by Grant-in-Aid for Young Scientists (B) 17740106 and 20740101, Grant-in-Aid for Scientific Research (C) 23540263, Grant-in-Aid for Scientific Research on Innovative Areas 18H05442 from the Japan Society for the Promotion of Science (JSPS), and by Satellite Research in 2017-2020 from Astrobiology Center, NINS. H.I. was supported by JSPS KAKENHI Grant Numbers JP16H02169, JP23244038.

This research has made use of the NASA Exoplanet Archive, which is operated by the California Institute of Technology, under contract with the National Aeronautics and Space Administration under the Exoplanet Exploration Program.
This research has made use of data obtained from or tools provided by the portal \href{http://exoplanet.eu/catalog/}{exoplanet.eu} of The Extrasolar Planets Encyclopaedia. 
This research has made use of the SIMBAD database, operated at CDS, Strasbourg, France \citep{Wenger2000}.
This work presents results from the European Space Agency (ESA) space mission Gaia. Gaia data are being processed by the Gaia Data Processing and Analysis Consortium (DPAC). Funding for the DPAC is provided by national institutions, in particular the institutions participating in the Gaia MultiLateral Agreement (MLA). The Gaia mission website is \href{https://www.cosmos.esa.int/gaia}{https://www.cosmos.esa.int/gaia}. The Gaia archive website is \href{https://archives.esac.esa.int/gaia}{https://archives.esac.esa.int/gaia}.

%



\software{\texttt{astropy} \citep{2013A&A...558A..33A,2018AJ....156..123A}, 
\texttt{Numpy} \citep{Harris2020}, 
\texttt{Scipy} \citep{Virtanen2020}, 
\texttt{pandas} \citep{Reback2022}, \texttt{matplotlib} \citep{Hunter2007}, \texttt{isochrones} \citep{Morton2015}, \texttt{PyMultiNest} \citep{Feroz2019}, 
\texttt{RadVel} \citep{Fulton2018}, 
\texttt{htof} \citep{GMBrandt2021}, 
\texttt{orvara} \citep{Brandt2021b}
}



\appendix
\section{The additional figures and tables}
\label{sec:appendix_A}

\begin{figure*}[ht]
\centering
\includegraphics[width=0.48\linewidth]{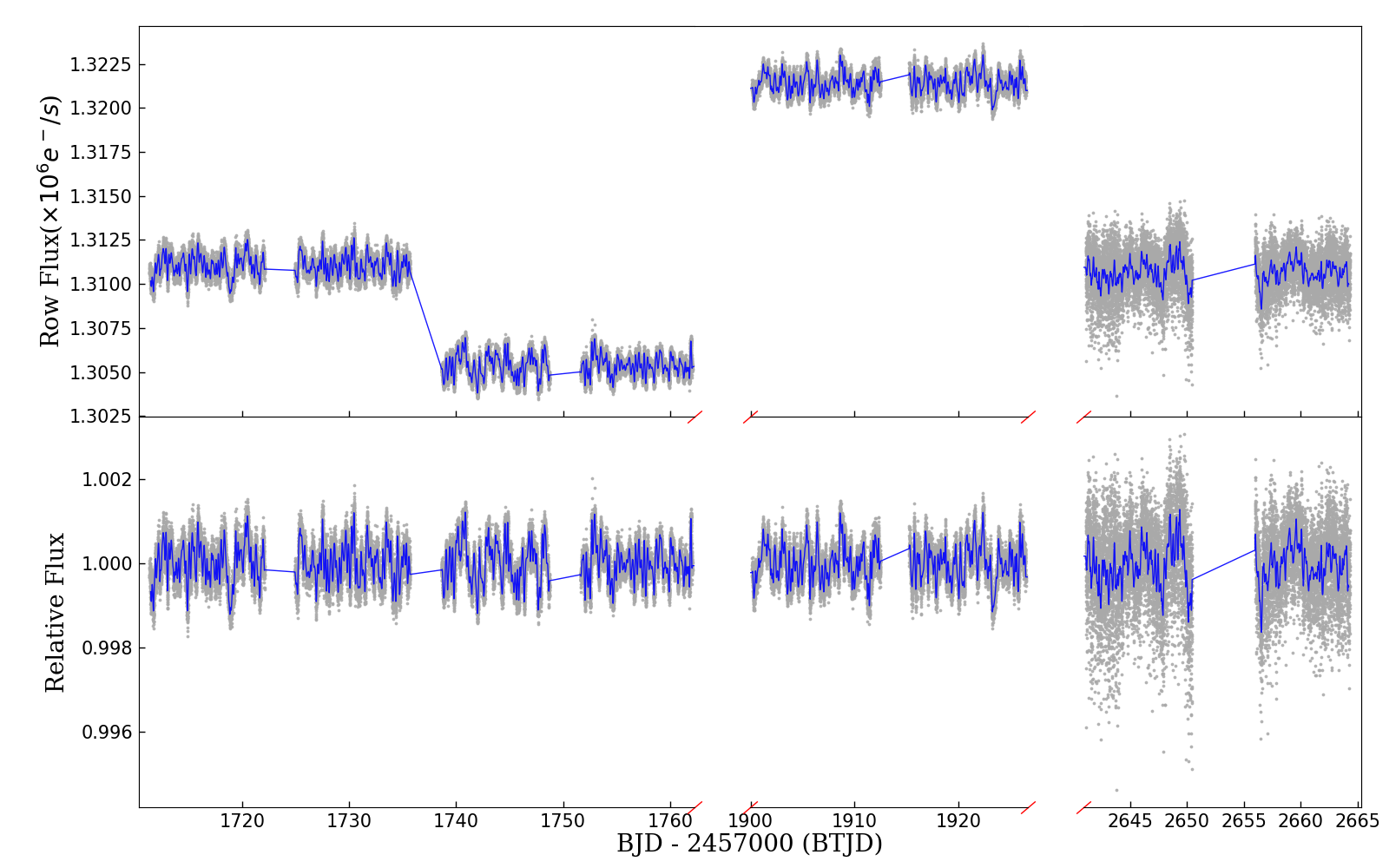}
\includegraphics[width=0.48\linewidth]{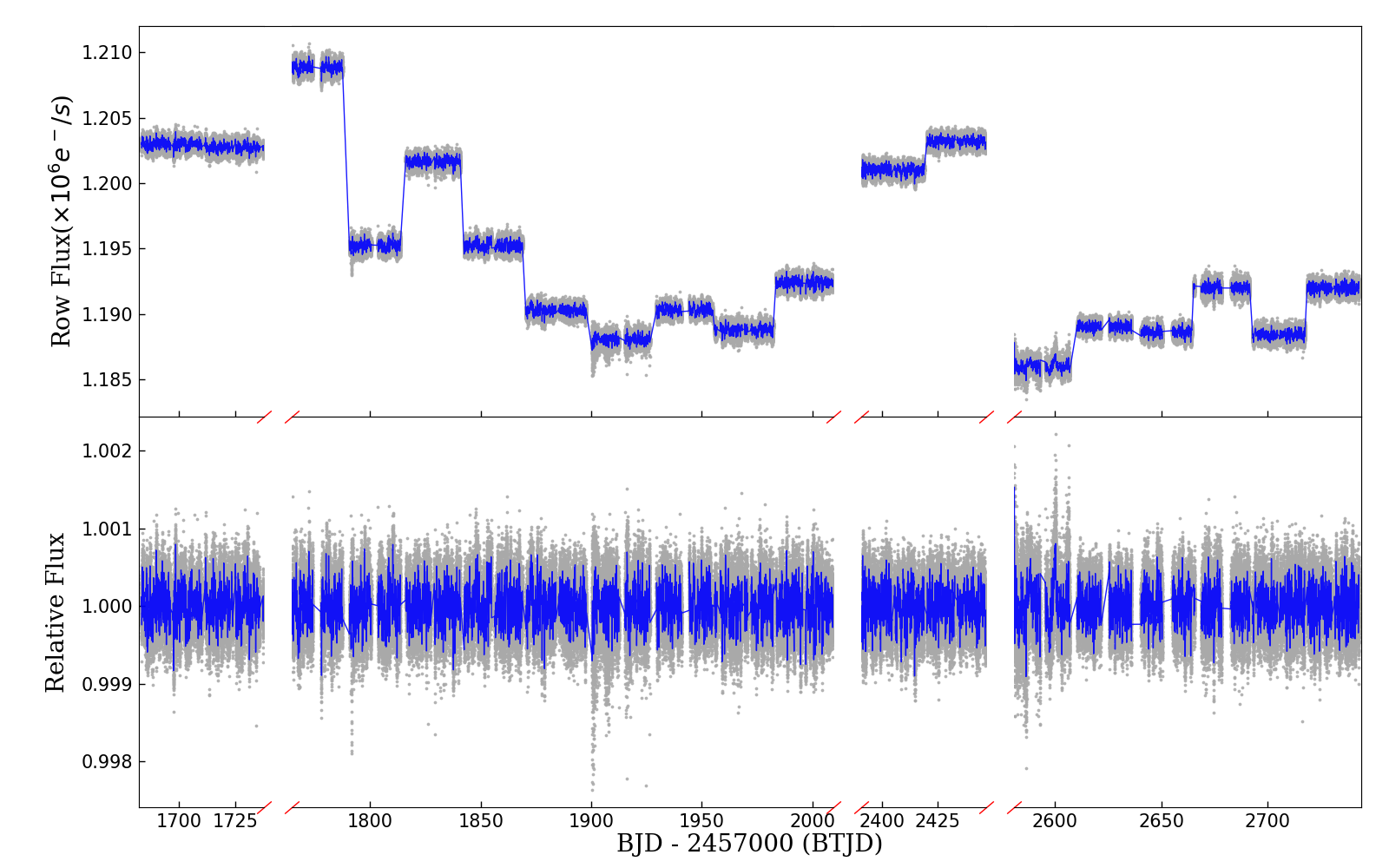}
\caption{Light curves of HD\,112570 (left panel) and HD\,154391 (right panel) generated by the TESS SPOC pipeline. Each panel shows the raw (top) and corrected (bottom) 2-minute cadence light curves. Additionally, a smoothed version of the light curve, obtained using a 0.1-day boxcar filter, is depicted in each subpanel with a blue curve.\label{fig:twotargetlc}}
\end{figure*}

\begin{figure*}[ht]
\centering
\includegraphics[width=0.48\linewidth]{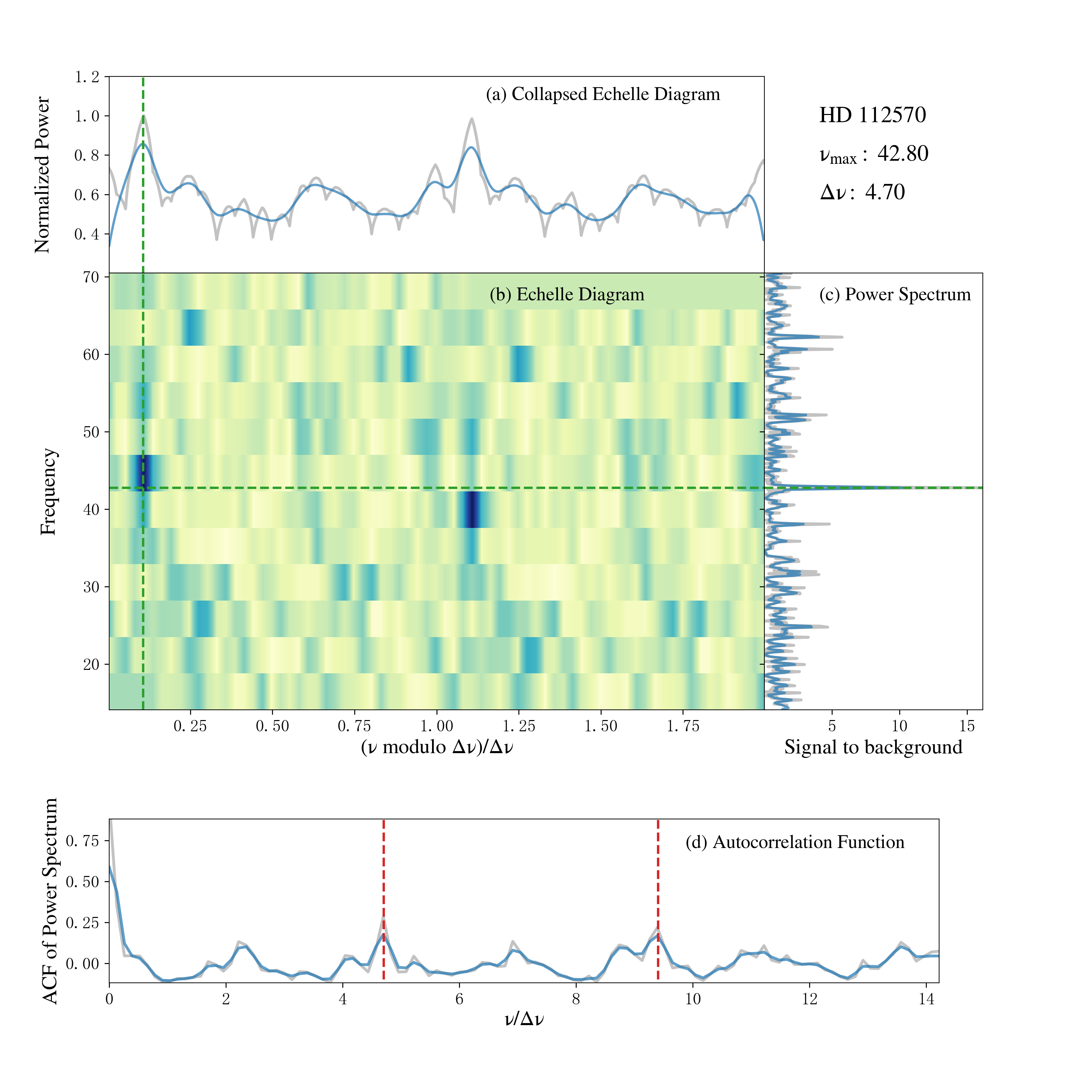}
\includegraphics[width=0.48\linewidth]{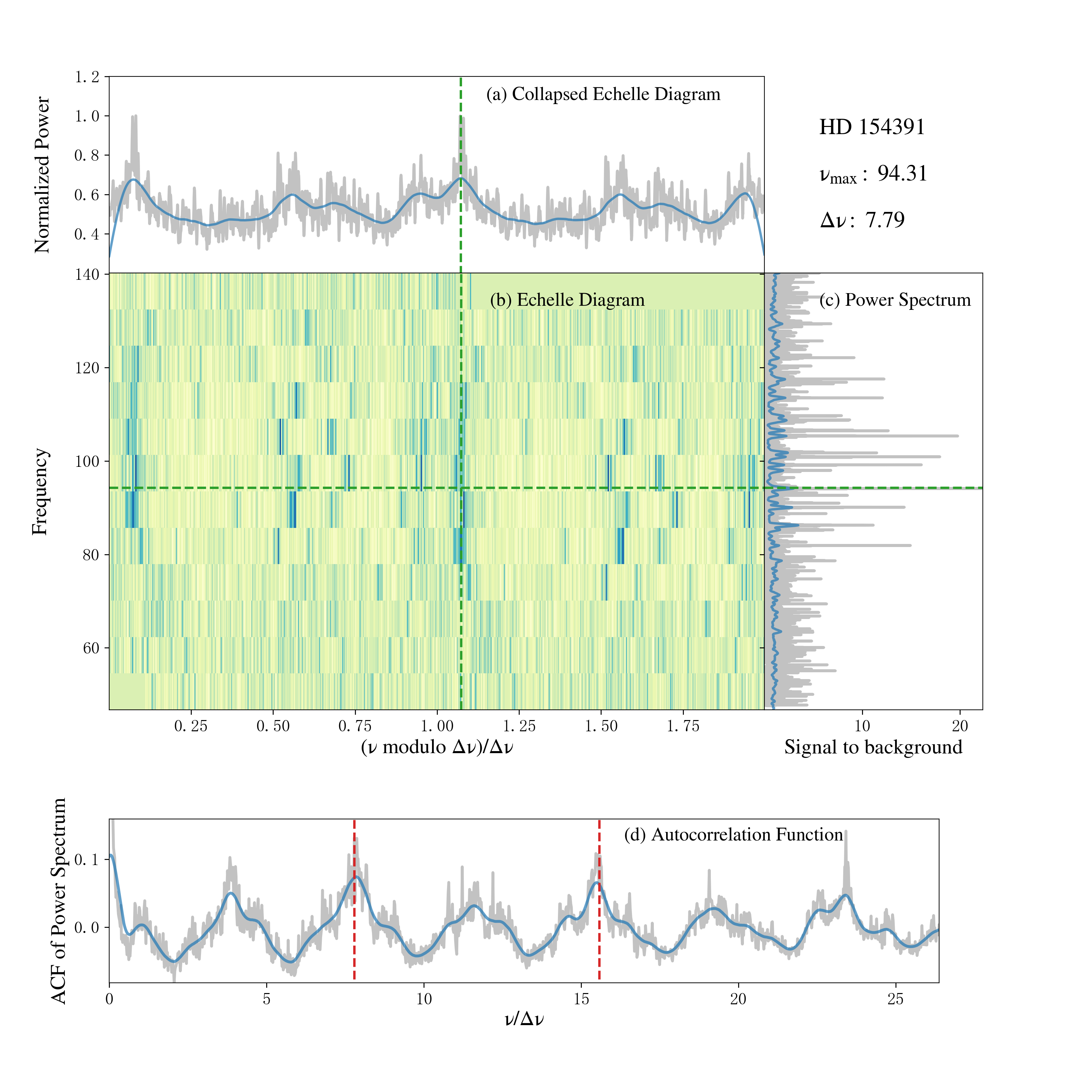}
\caption{The diagnostic plot of HD\,112570 (left panel) and HD\,154391 (right panel).The diagram includes:(a) The collapsed \'echelle diagram.(b) The \'echelle diagram.(c) The power spectrum, selected as a range of 3$\Delta\nu$ around $\nu_{\rm{max}}$ after subtracting the background. (d) The autocorrelation function resulting from autocorrelation function (ACF) analysis of the power spectrum. In both (a) and (b), the green dashed line indicates the radial mode ridge. In (b) and (c), the green dashed line represents $\nu_{\rm{max}}$. In (d), the two red dashed lines correspond to the positions of $\Delta\nu$ and twice $\Delta\nu$, from left to right.\label{fig:Deltanu_targets}}
\end{figure*}

\begin{figure*}[ht]
\plotone{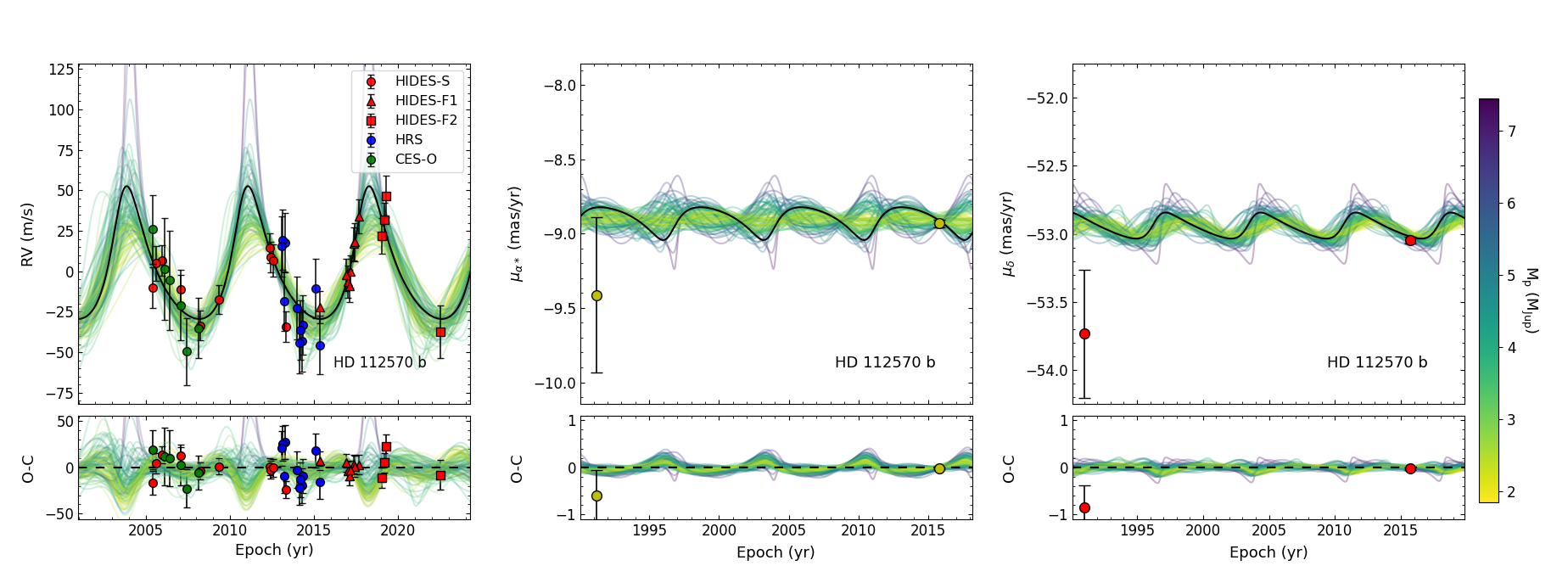}
\caption{Left panel: radial velocity curve of HD\,112570. The red points represent the data from HIDES. The blue and green points represent the data from CES-O and HRS, respectively. Middle and right panel: astrometric acceleration in right ascension and declination. The points near epoch 1991 are measured from Hipparcos, and the points near epoch 2016 are from Gaia EDR3. The black lines represent the best-fit orbit, and the colored lines, color-coded by the companion's mass, indicate the possible orbital solution randomly drawn from the MCMC chain. All figures are post-processed with \texttt{orvara}.
\label{fig:rv_pm_112570}}
\end{figure*}

\begin{figure*}[ht]
\plotone{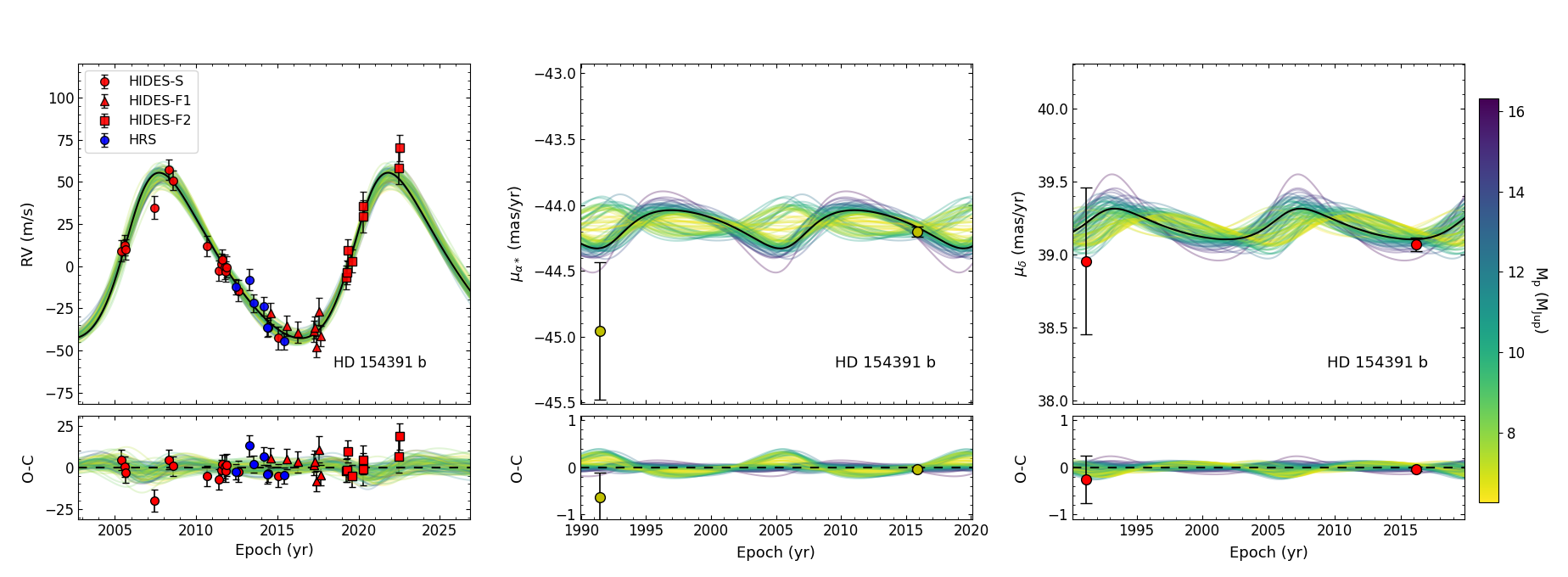}
\caption{Radial velocity curve and astrometric acceleration of HD\,154391. Same as Figure \ref{fig:rv_pm_112570}.
\label{fig:rv_pm_154391}}
\end{figure*}

\begin{figure}[!]
\plotone{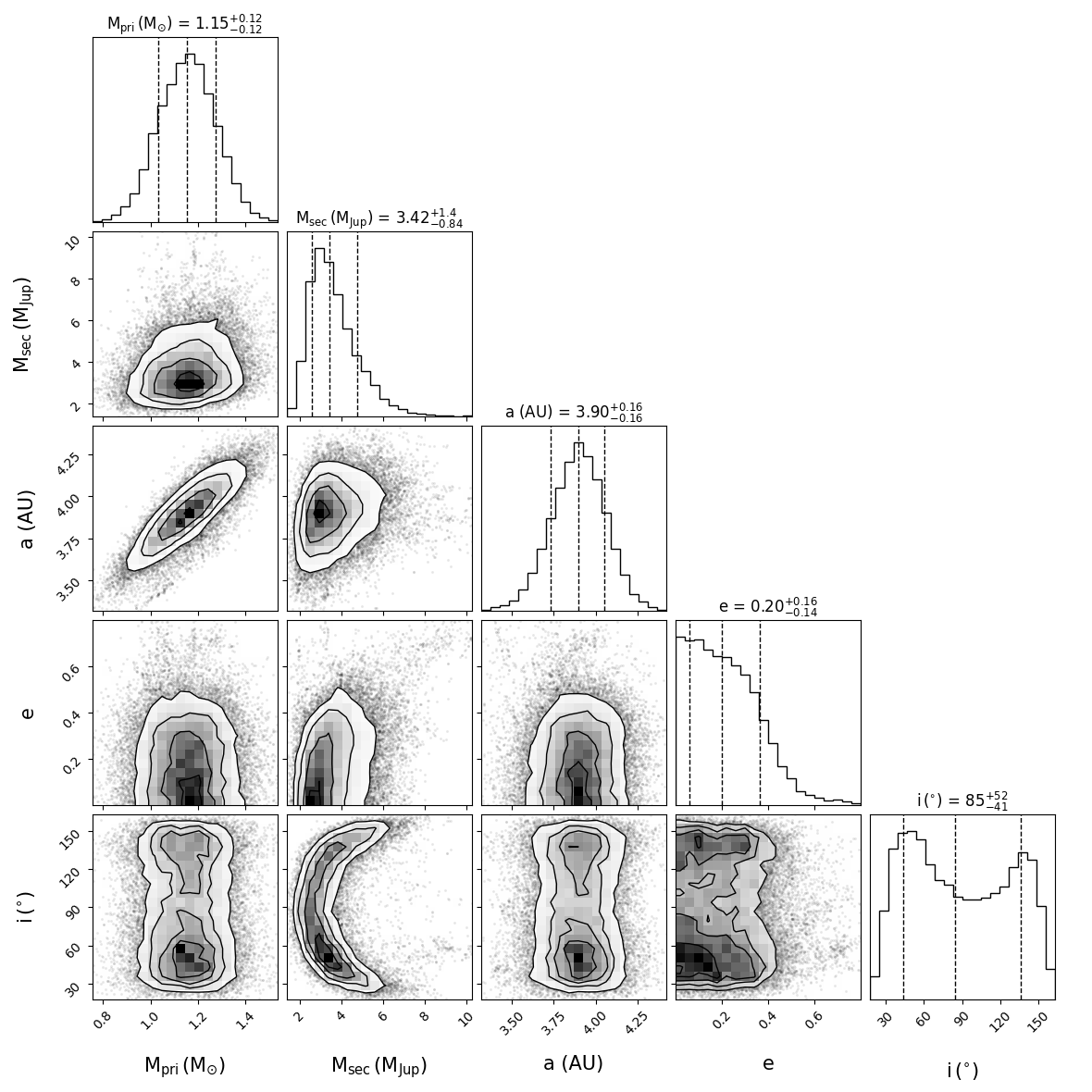}
\caption{Corner plot for HD\,112570. \texttt{orvara} posteriors for primary mass $M_{\rm pri}$, secondary mass $M_{\rm sec}$, semi-major axis $a$, eccentricity $e$, and inclination $i$.
\label{fig:corner_112570}}
\end{figure}

\begin{figure}[!]
\plotone{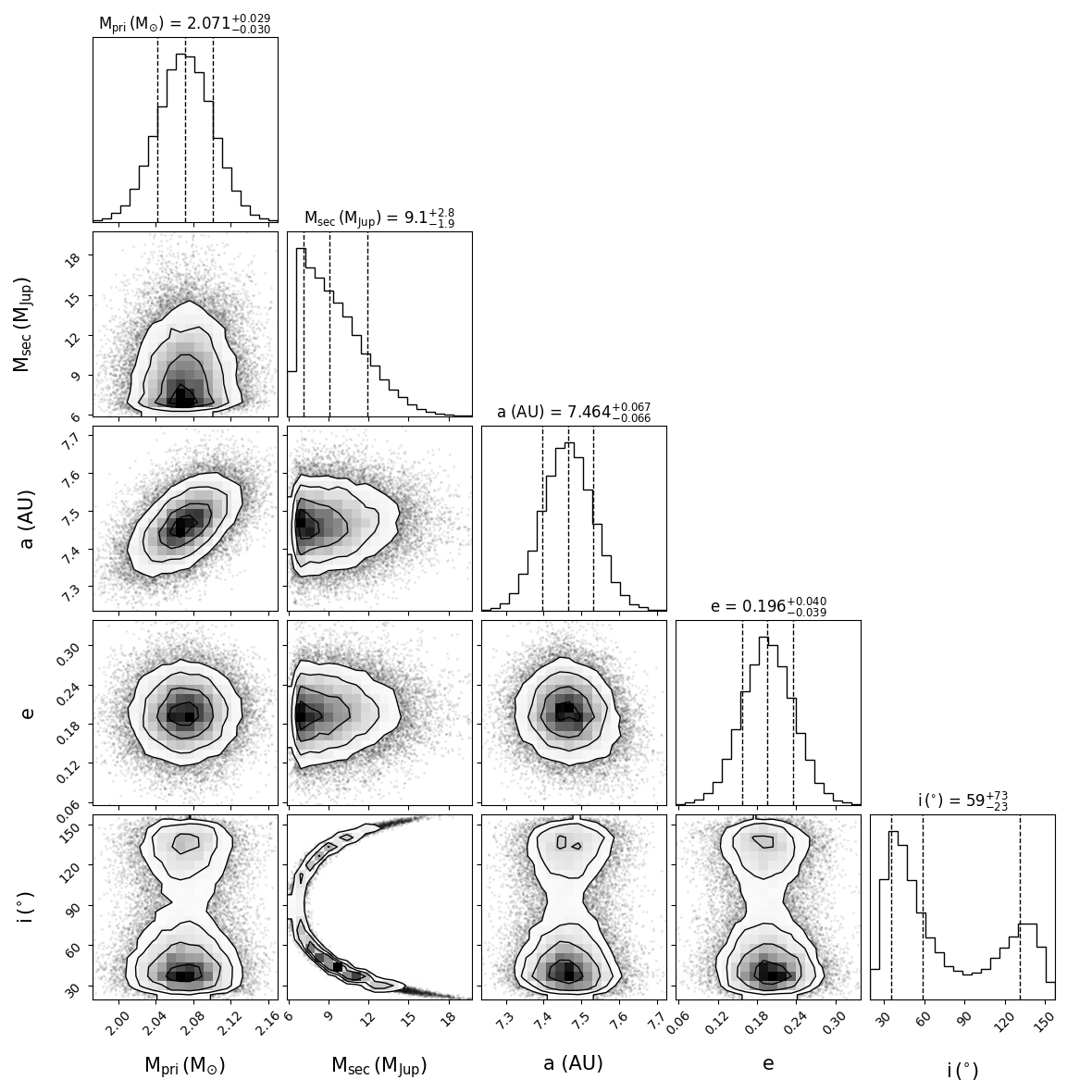}
\caption{Corner plot for HD\,154391. \texttt{orvara} posteriors for primary mass $M_{\rm pri}$, secondary mass $M_{\rm sec}$, semi-major axis $a$, eccentricity $e$, and inclination $i$.
\label{fig:corner_154391}}
\end{figure}

\begin{table*}[ht]
\caption{HGCA Astrometry}\label{Tab:hgca}
\begin{tabular*}{\textwidth}{@{}@{\extracolsep{\fill}}lccccccccc@{}}
\hline \hline
 Name & Data & $\mu_{\alpha \star}$& $\sigma[\mu_{\alpha \star}]$&$\mu_{\delta}$& $\sigma[\mu_{\delta}]$&  Correlation& Epoch, $\alpha$ & Epoch, $\delta$ & $\chi^{2}$\\
 & Source & \multicolumn{2}{c}{$({\rm mas\,yr}^{-1})$} &\multicolumn{2}{c}{$({\rm mas\,yr^{-1})}$} & Coefficient& \multicolumn{2}{c}{(yr)}\\
\hline
HD\,112570&Hip&$-9.412$&0.524&$-53.733$&$0.471$&0.344&1991.19&1991.04\\
&Hip-Gaia&$-8.896$&0.014&$-52.944$&0.014&0.489&&\\
&Gaia&$-8.931$&0.024&$-53.046$&0.027&0.512&2015.78&2015.73&11.5\\
HD\,154391&Hip&$-44.956$&0.519&$38.958$&$0.502$&0.051&1991.44&1991.19\\
&Hip-Gaia&$-44.140$&0.015&$39.203$&0.014&0.019&&\\
&Gaia&$-44.202$&0.038&$39.073$&0.046&0.089&2015.85&2016.18&9.0\\
\hline
\end{tabular*}
\tablecomments{The third to sixth columns represent proper motions and associated uncertainties. The last column lists the $\chi^2$ value for a constant proper motion model with two degrees of freedom \citep{Brandt2021}.}
\end{table*}

\begin{table*}[!]
\centering
\caption{Radial Velocities for HD\,112570}\label{Tab:RV_112570}
\begin{tabular*}{\textwidth}{@{}@{\extracolsep{\fill}}lccccccc@{}}
\hline \hline
 $\rm JD-2450000$ & RV ($\rm m\,s^{-1}$) & Error ($\rm m\,s^{-1}$) & Instrument&$\rm JD-2450000$ & RV ($\rm m\,s^{-1}$) & Error ($\rm m\,s^{-1}$) & Instrument\\
\hline
3514.1271 & 166.76 & 21.32 & CES-O & 4126.3374 & $-12.62$ & 3.87 & HIDES-S \\
3776.2802 & 142.54 & 31.43 & CES-O &4558.0089 & $-34.99$ & 3.94 & HIDES-S \\
3891.0835 & 135.37 & 30.57 & CES-O &4954.1124 & $-19.03$ & 3.74 & HIDES-S \\
4131.3816 & 119.91 & 21.69 & CES-O &6064.0347 & 12.87 & 4.16 & HIDES-S \\
4257.0930 & 91.32 & 20.68 & CES-O &6084.9958 & 7.43 & 4.89 & HIDES-S \\
4520.3169 & 105.77 & 18.55 & CES-O &6139.0134 & 5.18 & 5.63 & HIDES-S \\
6318.3497 & 36.62 & 5.62 & HRS &6414.1305 & $-35.71$ & 4.56 & HIDES-S \\
6344.2550 & 40.53 & 5.80 & HRS &7142.0657 & $-23.69$ & 5.74 & HIDES-F1 \\
6384.1575 & 2.74 & 4.38 & HRS &7725.3294 & $-3.87$ & 5.63 & HIDES-F1 \\
6404.1169 & 38.72 & 4.71 & HRS &7763.2388 & $-8.11$ & 5.80 & HIDES-F1 \\
6645.3904 & $-1.55$ & 7.87 & HRS &7791.2577 & $-10.54$ & 5.57 & HIDES-F1 \\
6700.3405 & $-23.00$ & 6.03 & HRS &7809.2527 & $-1.59$ & 5.86 & HIDES-F1 \\
6729.2150 & $-15.12$ & 4.83 & HRS &7885.1893 & 15.62 & 6.23 & HIDES-F1 \\
6760.1616 & $-22.04$ & 6.33 & HRS &7907.9676 & 16.67 & 8.45 & HIDES-F1 \\
6791.0737 & $-12.29$ & 4.85 & HRS &7992.9534 & 32.44 & 6.90 & HIDES-F1 \\
7058.3529 & 10.37 & 4.40 & HRS &8490.3892 & 20.55 & 7.50 & HIDES-F2 \\
7148.1142 & $-24.42$ & 3.36 & HRS &8551.0986 & 30.33 & 6.83 & HIDES-F2 \\
3518.9926 & $-11.66$ & 9.70 & HIDES-S &8584.3179 & 44.84 & 10.12 & HIDES-F2 \\
3599.9826 & 3.45 & 6.88 & HIDES-S &9760.1027 & $-38.85$ & 14.02 & HIDES-F2 \\
3719.3592 & 5.17 & 4.80 & HIDES-S \\
\hline
\end{tabular*}
\tablecomments{The RV offsets between HIDES-S, -F1 and -F2 are fixed to 0 for using the same reference spectra of each mode.}
\end{table*}

\begin{table*}[!]
\centering
\caption{Radial Velocities for HD\,154391}\label{Tab:RV_154391}
\begin{tabular*}{\textwidth}{@{}@{\extracolsep{\fill}}lccccccc@{}}
\hline \hline
 $\rm JD-2450000$ & RV ($\rm m\,s^{-1}$) & Error ($\rm m\,s^{-1}$) & Instrument& $\rm JD-2450000$ & RV ($\rm m\,s^{-1}$) & Error ($\rm m\,s^{-1}$) & Instrument\\
\hline
6083.1749 & 14.41 & 2.46 & HRS &5881.8687 & $-3.77$ & 3.61 & HIDES-S \\
6404.3258 & 18.84 & 4.95 & HRS &6142.0053 & $-17.76$ & 3.72 & HIDES-S \\
6493.1287 & 4.82 & 3.43 & HRS &7047.3806 & $-45.91$ & 4.35 & HIDES-S \\
6730.3272 & 3.01 & 3.82 & HRS &6864.0994 & $-31.32$ & 2.98 & HIDES-F1 \\
6792.2636 & $-9.70$ & 4.14 & HRS &7238.0263 & $-38.73$ & 3.13 & HIDES-F1 \\
6820.2156 & $-9.72$ & 3.04 & HRS &7476.2591 & $-42.52$ & 3.59 & HIDES-F1 \\
7174.2134 & $-17.78$ & 3.09 & HRS &7856.2798 & $-41.80$ & 3.35 & HIDES-F1 \\
3521.1921 & 5.89 & 3.39 & HIDES-S &7857.3065 & $-39.73$ & 3.81 & HIDES-F1 \\
3599.1055 & 9.00 & 3.03 & HIDES-S &7895.1900 & $-51.00$ & 3.03 & HIDES-F1 \\
3615.0588 & 6.65 & 2.97 & HIDES-S &7957.9905 & $-30.24$ & 6.20 & HIDES-F1 \\
4258.2611 & 31.46 & 4.20 & HIDES-S &7991.9998 & $-44.75$ & 3.24 & HIDES-F1 \\
4589.2290 & 53.94 & 2.98 & HIDES-S &8562.3056 & $-9.89$ & 4.59 & HIDES-F2 \\
4672.9921 & 47.56 & 2.58 & HIDES-S &8592.2885 & $-6.90$ & 4.59 & HIDES-F2 \\
5437.9606 & 8.55 & 2.99 & HIDES-S &8614.2593 & 6.28 & 3.84 & HIDES-F2 \\
5718.0816 & $-5.72$ & 2.93 & HIDES-S &8706.1368 & $-0.24$ & 4.11 & HIDES-F2 \\
5766.1157 & $-1.94$ & 3.14 & HIDES-S &8956.1618 & 26.27 & 8.05 & HIDES-F2 \\
5784.9848 & 0.50 & 2.97 & HIDES-S &8960.3096 & 32.45 & 6.83 & HIDES-F2 \\
5850.9279 & $-3.70$ & 4.86 & HIDES-S &9760.0144 & 54.97 & 8.06 & HIDES-F2 \\
5853.8910 & $-6.40$ & 3.20 & HIDES-S &9767.0684 & 66.96 & 5.71 & HIDES-F2 \\
\hline
\end{tabular*}
\tablecomments{The RV offsets between HIDES-S, -F1 and -F2 are fixed to 0 for using the same reference spectra of each mode.}
\end{table*}


\bibliography{sample631}{}
\bibliographystyle{aasjournal}



\end{document}